\documentclass[preprint,12pt,endfloats*,prb]{revtex4}
\usepackage{natbib}
\usepackage[latin1]{inputenc}
\usepackage[T1]{fontenc}
\usepackage{amsfonts}
\usepackage{amssymb}
\usepackage{amsmath}
\usepackage{wasysym}
\usepackage{enumitem}
\usepackage{graphicx}
\usepackage{color}
\usepackage{euscript}
\definecolor{rouge}{rgb}{0.8,0.19,0.11}
\definecolor{bleu}{rgb}{0.,0.,0.5}
\def\u_vu{\tilde{u}}
\def\lrc#1{\left\langle #1\right\rangle}
\def\lrp#1{\left( #1\right)}
\def\lrb#1{\left[ #1\right]}

\def\derivp#1#2{\frac{\partial #1}{\partial #2}}
\def\derivt#1#2{\frac{d #1}{d #2}}
\def\derivtmp#1#2{\frac{\overline{D}_p #1}{D #2}}
\def\eref#1{Eq.~(\ref{#1})}
\graphicspath{{Figures_eps/}{Figures/}}
\newlength{\myfigwidth}
\makeatletter
\def\@dotsep{1024}

\begin{document}


\title{Simulation of a particle-laden turbulent channel flow using an improved stochastic Lagrangian model}

\author{Boris Arcen}
 \email{boris.arcen@esstin.uhp-nancy.fr}
\author{Anne Tani\`ere}%
 \email{anne.taniere@esstin.uhp-nancy.fr}
\affiliation{%
LEMTA, Nancy-University, CNRS, ESSTIN, 54519 Vandoeuvre-l\`es-Nancy - France.
}%
\date{\today}
%
\begin{abstract}
The purpose of this paper is to examine the Lagrangian stochastic modeling of the fluid velocity seen by inertial particles in a non-homogeneous
turbulent flow. A new Langevin-type model, compatible with the transport equation of the drift velocity in the limits of low and high particle
inertia, is derived. It is also shown that some previously proposed stochastic models are not compatible with this transport equation in the
limit of high particle inertia. The drift and diffusion parameters of these stochastic differential equations are then estimated using DNS data.
It is observed that, contrary to the conventional modeling, they are highly space-dependent and anisotropic. To investigate the performance of
the present stochastic model, a comparison is made with DNS data as well as with  two different stochastic models. A good prediction of the
first and second order statistical moments of the particle and fluid seen velocities is obtained with the three models considered. Even for some
components of the triple particle velocity correlations, an acceptable accordance is noticed. The performance of the three different models
mainly diverges for the particle concentration and the drift velocity. The proposed model is seen to be the only one which succeeds in
predicting the good evolution of these latter statistical quantities for the range of particle inertia studied.
%
\end{abstract}




\maketitle

%
%
%
%
%
\section{Introduction}
In the past years, several numerical methods have been employed to study the dispersion of solid particles in turbulent flows. Generally, small
enough particles are considered in order to treat them as point-particles.\cite{mari,ReeksJFM,mcla89,rouson} Assuming that the drag force is
only of importance, the link between the motion of an inertial particle and the carrier fluid is given by the following system of equations :
\begin{align}
&\frac{dx_{p,i}}{dt}= v_{p,i}, \label{equa_evol_position_particule}\\
&\frac{dv_{p,i}}{dt}= \frac{\u_vu_i-v_{p,i}}{\tau_p}\,, \label{equa_evol_vitesse_particule}
\end{align}
where $x_{p,i}$ and $v_{p,i}$ are the particle position and velocity, $\tau_p$ is the particle relaxation time which is expressed in terms of
the drag coefficient and of the magnitude of the relative velocity, and $\u_vu_i=u_i(\mathbf{x}_{p},t)$ is the fluid velocity at the particle
location.
Under these considerations, the main difficulty then lies in the proper computation of the fluid velocity at each particle location. The first
possibility is to use Direct Numerical Simulation (DNS).\cite{mcla89,ferelg_POF_03} This technique gives the best estimation of the fluid
velocity seen by particles. Nevertheless, it necessitates very high computational ressources. A more affordable numerical way is provided by
Large Eddy Simulation (LES).\cite{masaso_pof_2008} Contrary to DNS, a model which takes the residual fluid dynamic (i.e. at the sub-grid scale)
into account should be used to predict the instantaneous fluid velocity seen by particles. Finally, when the computational cost of this latter
technique is still too high, it is then possible to make use of macroscopic numerical simulation such as Reynolds-Averaged Navier-Stokes (RANS).
The use of a RANS-Lagrangian method to describe the motion of solid particles in a turbulent two-phase flow necessitates to generate the
fluctuating velocity of the carrier phase at particle location.\cite{maspan_review} In this framework, averaged quantities such as the mean
velocity and some of the mean turbulent characteristics of the carrier phase are determined by solving the Reynolds-Averaged Navier-Stokes
equations. The time integration of the equations governing the motion of inertial particles [Eqs.~(\ref{equa_evol_position_particule}) and
(\ref{equa_evol_vitesse_particule})] requires the knowledge of the instantaneous velocity of the fluid at the particle location. The
reconstruction of the random nature of the fluctuations along inertial particle trajectories can be achieved, for instance, using a stochastic
Lagrangian models.\cite{simonin93,oest_04,ilihan_2004,Reeks05,piasimvil_IJMF_2008,dehbi_IJMF_2008} Most of these models for the simulation of
turbulent two-phase flows involves specific formulations based on the Langevin model which can be written in a general form as :
\begin{equation}\label{eq:GLM_forme_generale}
d\u_vu_i=A_i dt+ B_{ij} dW_j\,,
\end{equation}
for the instantaneous fluid velocity at particle location. In this latter stochastic differential equation (SDE), $A_i$ is the drift vector,
$B_{ij}$ is the diffusion matrix, and $dW_j$ are the increments of a vector-valued Wiener process with independent components. Some important
properties of these increments are that they are non-differentiable and normally distributed with mean $\lrc{W_i(t+dt)-W_j(t)}=0$ and variance
$\lrc{\lrb{W_i(t+dt)-W_j(t)}^2}=dt\,\delta_{ij}$.\cite{gardin} In order to predict the fluid velocity seen by inertial particles, one has to
model the drift vector and the diffusion matrix. In the present study, we focus on the models for the drift vector proposed by \citet{simonin93}
 \begin{equation}\label{eq:GLM_simonin_inst}
A_i=-\frac{1}{\rho_f}\derivp{\lrc{p}}{x_i}+\nu \derivp{^2\lrc{u_i}}{x_j\partial
x_j}+\lrp{v_{p,j}-\u_vu_j}\derivp{\lrc{u_i}}{x_j}+G_{ij}\lrp{\u_vu_j-\lrc{u_j}}\,,
\end{equation}
and by \citet{minpei_01}
 \begin{equation}\label{eq:GLM_minier_inst}
A_i=-\frac{1}{\rho_f}\derivp{\lrc{p}}{x_i}+\nu \derivp{^2\lrc{u_i}}{x_j\partial
x_j}+\lrp{\lrc{v_{p,j}}-\lrc{\u_vu_j}}\derivp{\lrc{u_i}}{x_j}+G_{ij}\lrp{\u_vu_j-\lrc{u_j}}\,.
\end{equation}
In these expressions, $\nu$ is the kinematic viscosity, $\rho_f$ is the fluid density, $p$ stands for the pressure, $u_i$ is the fluid velocity,
$v_{p,i}$ is the solid particle velocity and $G_{ij}$ is the drift matrix. The difference between both models lies in the form of the third term
which describes the crossing-trajectory effect.\cite{csanad} In the model proposed by \citet{simonin93}, this term is written as a function of
the instantaneous relative velocity between the particle and the fluid while \citet{minpei_01} suggested to express it as a function of the mean
relative velocity.\\
It has to be noted that in the limit of low particle inertia ($\tau_p\ll 1$), both models give the well-known Generalized Langevin Model (GLM)
derived by \citet{pope_83} to predict the motion of fluid particle in a turbulent flow. Nevertheless, as it will be shown in Sec. II, these two
previous stochastic models are not compatible with the transport equation of the drift velocity (mean fluctuating fluid velocity at particle
location) for large particle inertia. In order to correct this discrepancy, a new form of the drift vector is proposed. In Sec. III, the method
used to derive the drift and diffusion parameters of these stochastic models is described, and the estimated values obtained using DNS data are
presented. The performance of the proposed functional form of the drift vector is then assessed by comparison with DNS data in Sec. IV. Finally,
concluding remarks are given in the last section.

%
%
%
\section{Exact and modeled transport equations of the drift velocity}
In this section, we study the Langevin models proposed by \citet{simonin93} and \citet{minpei_01} through the transport equation of the drift velocity in the limit of low and high particle inertia. Based on this study, a new model for the drift vector $A_i$ is proposed.
%
\subsection{Degenerate equations for low and high particle inertia} 
Let us consider the gas-solid flow from an Eulerian (macroscopic) point of view. The exact transport equation for the statistical moments of the particle and fluid seen velocities, as well as for the fluid seen-particle velocity correlations, can be derived, for example, from the transport equation of a joint probability density function for the particle and fluid seen velocities.\cite{simo_ivk,minpei_01,panmas,Reeks05} The exact transport equation of the drift velocity, $\lrc{\u_vu'_i}=\lrc{\u_vu_i-\lrc{u_i}}$, can be written as \cite{minpei_01}
\begin{align}\label{eq:transp_derive_exacte}
\alpha_p\rho_p\derivtmp{}{t}\lrc{\u_vu'_i}=&\alpha_p\rho_p\lrb{\derivp{}{x_j}\lrp{\lrc{u'_iu'_j}-\lrc{\u_vu'_iv'_{p,j}}}}
-\lrc{\u_vu'_iv'_{p,j}}\derivp{}{x_j}\lrp{\alpha_p\rho_p}\nonumber\\
&-\alpha_p\rho_p\lrp{\lrc{v_{p,j}}-\lrc{\u_vu_j}}\derivp{\lrc{u_i}}{x_j}-\alpha_p\rho_p\lrc{\u_vu'_j}\derivp{\lrc{u_i}}{x_j}\nonumber\\
&-\alpha_p\rho_p\lrp{-\frac{1}{\rho_f}\derivp{\lrc{p}}{x_i}+\nu \derivp{^2\lrc{u_i}}{x_j\partial x_j}}+ \alpha_p\rho_p\lrc{\derivt{\u_vu_i}{t}}
\,,
\end{align}
where $\overline{D}_p(\cdot)/Dt=\partial(\cdot)/\partial t + \lrc{v_{p,j}}\partial(\cdot) /\partial x_j$, and $\alpha_p$ is the particle volume fraction.\\
In the limit of vanishing particle inertia ($\tau_p\ll 1$), a solid particle behaves like a fluid particle tracer. Its velocity is equal to that
of a fluid particle, the statistical moments of the fluid and particle velocities are thus identical. Moreover, the drift velocity is zero since
this kind of particles samples homogeneously the turbulent flow field and the particle volume fraction is constant if the particles are
uniformly distributed initially. As a consequence, it can be found, from equation~(\ref{eq:transp_derive_exacte}), that the average of the time
variation of the fluid velocity seen becomes equal to
\begin{align}\label{eq:transp_derive_exacte_petit_taup}
\lrc{\derivt{\u_vu_i}{t}}=-\frac{1}{\rho_f}\derivp{\lrc{p}}{x_i}+\nu \derivp{^2\lrc{u_i}}{x_j\partial x_j}\,.
\end{align}
The averaged Navier-Stokes equations are thus recovered.\\
The opposite limit case, i.e. high particle inertia ($\tau_p\gg 1$), is also of great importance when studying gas-solid flows since the
trajectories of such particles become completely independent of the fluid motion. In such a case, the particle velocity remains nearly identical
to its initial value, the fluid seen-particle velocity correlations, the second and higher statistical moments of the particle velocity as well
as the drift velocity tend to zero. Moreover, the particle volume fraction keeps a constant value across the fluid flow if particles are
initially distributed uniformly. In the present study and without loss of generality, the velocity of these high inertia particles is considered
identical to the mean fluid velocity. Under these considerations, it can be found from equation~(\ref{eq:transp_derive_exacte}) that
\begin{align}\label{eq:transp_derive_exacte_grand_taup}
\lrc{\derivt{\u_vu_i}{t}}=-\frac{1}{\rho_f}\derivp{\lrc{p}}{x_i}+\nu \derivp{^2\lrc{u_i}}{x_j\partial x_j}-\derivp{}{x_j}\lrc{u'_iu'_j}\,.
\end{align}
These two previous equations give the asymptotic limits of the average time derivative of the fluid velocity seen by particles. Besides, they can be used in order to verify that the Langevin models generally used to predict the evolution in time of this fluid velocity are correct in the limits of low and high particle inertia.\\
For example, let us consider first the model of the drift vector proposed by \citet{simonin93}, i.e. equation~(\ref{eq:GLM_simonin_inst}). The average of equation~(\ref{eq:GLM_forme_generale}) in the limit of low particle inertia yields
\begin{align}
\lrc{\derivt{\u_vu_i}{t}}=-\frac{1}{\rho_f}\derivp{\lrc{p}}{x_i}+\nu \derivp{^2\lrc{u_i}}{x_j\partial x_j}\,,
\end{align}
since $\lrc{B_{ij}dW_j} = 0$. Thus, the model of the drift vector by \citet{simonin93} is able to produce the correct limit of the mean time
increment of the fluid velocity seen in this particular case. When the particle inertia becomes high, the same expression is obtained from this
model. In comparison with the exact one given by equation~(\ref{eq:transp_derive_exacte_grand_taup}), it is noticed that the divergence of the
Reynolds stress tensor is missing. Therefore, we can expect that some discrepancies could occur in the prediction of the momentum exchange
between the dispersed and carrier phases for high particle inertia. Considering the model proposed by \citet{minpei_01}, it can be seen that the
expressions of the mean time increment of the fluid velocity seen in the limits of low and high particle inertia are identical to those obtained
with the model of \citet{simonin93} This model is thus also not compatible with the transport equation of the drift velocity in the limit of
high particle inertia. In the next section, a new model of the drift vector which makes possible the prediction of the theoretical limits given
above is proposed.
%
%
\subsection{Proposal of a new model}
From the conclusions drawn in the previous section, we have designed a new model which gives the proper limits of the drift velocity transport
equations in the cases of low and high particle inertia. This model for the drift vector $A_i$ is
\begin{equation}\label{eq:GLM_nous_inst}
A_i=-\frac{1}{\rho_f}\derivp{\lrc{p}}{x_i}+\nu \derivp{^2\lrc{u_i}}{x_j\partial
x_j}+\lrp{v_{p,j}-\u_vu_j}\derivp{\lrc{u_i}}{x_j}+G_{ij}\lrp{\u_vu_j-\lrc{u_j}}+\derivp{}{x_k}\lrp{\lrc{\u_vu'_iv'_{p,k}}-\lrc{u'_iu'_k}}\,.
\end{equation}
This drift vector is mainly different from those proposed by \citet{simonin93} and \citet{minpei_01} due to the presence of the term
$\partial\lrp{\lrc{\u_vu'_iv'_{p,k}}-\lrc{u'_iu'_k}}/\partial x_k$ (i.e. the divergence of the difference between the fluid-particle covariances
and the Reynolds stresses).

First, it has to be noted that in the limit of low particle inertia the model proposed is also identical to the GLM derived by \citet{pope_83}
because $\lrc{\u_vu'_iv'_{p,k}}\rightarrow\lrc{u'_iu'_k\vphantom{\u_vu'_iv'_{p,k}}}$. In addition, the modeled averaged time increment of the
fluid velocity seen using \eref{eq:GLM_nous_inst} has the proper limits since
\begin{align}
\lrc{\derivt{\u_vu_i}{t}}=-\frac{1}{\rho_f}\derivp{\lrc{p}}{x_i}+\nu \derivp{^2\lrc{u_i}}{x_j\partial x_j}\,,
\end{align}
when $\tau_p\ll 1$, and
\begin{align}
\lrc{\derivt{\u_vu_i}{t}}=-\frac{1}{\rho_f}\derivp{\lrc{p}}{x_i}+\nu \derivp{^2\lrc{u_i}}{x_j\partial x_j}-\derivp{}{x_j}\lrc{u'_iu'_j}\,,
\end{align}
for $\tau_p\gg 1$.\\
The introduction of a supplementary term, which is a function of the Reynolds stresses, in the drift vector has been motivated by the necessity
for the stochastic model to be consistent with the transport equation of the drift velocity in the limit of high particle inertia. Nevertheless,
this additional term had to vanish in limit of low particle inertia in order to keep the model similar to the GLM. This has naturally led us to
add the fluid-particle covariance tensor which tends to the Reynolds stresses when $\tau_p\ll 1$ and to zero when $\tau_p\gg 1$. Moreover,
incorporating the proposed model in the transport equation of the drift equation, it can be seen that this new term moment is physically
consistent with the others.

At this point, we would like to emphasize the fact that the present model should be more suitable for predicting the fluid velocity seen by
large solid particles than the models proposed by \citet{simonin93} and \citet{minpei_01}, however, the presence of the fluid-particle
covariances in the expression increases the degree of complexity of the stochastic model.

It is also worth mentioning that the SDE for the fluctuating fluid velocity seen derived from this model (this SDE is presented hereafter)
presents similarities with the one proposed recently by \citet{bolo_IJMF_2006}. In this latter study, they concluded that a ``drift
correction'', which is a function of the fluid seen and particle velocities, should be included in the SDE in order to correctly predict the
concentration profiles of finite-inertia particles in a turbulent boundary layer. In fact, the last term in \eref{eq:GLM_nous_inst} plays this
role.

Before evaluating the performance of the proposed model, the procedure used to specify the parameters of the stochastic equation, i.e. the drift
and diffusion matrices ($G_{ij}$ and $B_{ij}$), is presented.
%
%
%
%
%
\section{Determination of the drift and diffusion matrices}
\subsection{Theoretical formalism and assumptions}
In order to test the capability of the proposed form for the drift vector to model the turbulence seen by inertial particles, the values of the components of the drift and diffusion matrices, $G_{ij}$ and $B_{ij}$, have to be specified. In stationary homogeneous isotropic turbulence and without a mean relative motion between the dispersed and carrier phases, the drift term is modeled has the inverse of the integral time scale of the fluid seen. The diffusion matrix is generally supposed independent of the particle inertia and is expressed as a function of the Kolmogorov's constant and dissipation rate of the mean turbulent kinetic energy according to the Kolmogorov similarity theory for the second-order Lagrangian velocity structure function in the inertial subrange.\cite{pope_87} It has to be noted that this model for the diffusion term is strictly valid in the limit of vanishing particle inertia and for high Reynolds number turbulent fluid flows.\\
In the case of non-homogeneous turbulence, the specification of the drift and diffusion matrices is even more complex. There are no models for these quantities which take properly the properties of such a turbulence into account. Consequently, we propose in the present study to determine $G_{ij}$ and $B_{ij}$ using data extracted from our channel flow DNS computation. A similar method to the one proposed in the study by \citet{pope_02}, which was devoted to the prediction of fluid particle trajectories in a turbulent homogeneous shear flow, is followed.\\
Firstly, in order to apply this method, the stochastic differential equation for the fluctuating fluid velocity at the solid particle location has to be derived from \eref{eq:GLM_forme_generale}. Since $d\u_vu'_i=d\u_vu_i-d\lrc{u_i}$, it can be shown that
 \begin{equation}\label{eq:GLM_simonin_fluc}
d\u_vu'_i=\widetilde{G}_{ij}\u_vu'_jdt + B_{ij} dW_j+\derivp{\lrc{u'_iu'_j}}{x_j}dt\,,
\end{equation}
when the model by \citet{simonin93} is used while
 \begin{equation}\label{eq:GLM_nous_fluc}
d\u_vu'_i=\widetilde{G}_{ij}\u_vu'_jdt + B_{ij} dW_j+\derivp{\lrc{\u_vu'_iv'_{p,j}}}{x_j}dt\,,
\end{equation}
with the present model for $A_i$ [\eref{eq:GLM_nous_inst}]. In these equations, $\widetilde{G}_{ij}=G_{ij}-\partial \lrc{u_i}/\partial x_j$. The
model of \citet{minpei_01} will not be considered in the rest of the present study for two reasons. This model suffers from the same drawback in
the limit of high particle inertia as the one suggested by \citet{simonin93}, consequently, only one of these models can be examined. In
addition, their model was designed to be used for the prediction of the instantaneous fluid velocity seen by particles and is thus not of
practical use for predicting the fluctuating part.\\
Secondly, since the method proposed by \citet{pope_02} is strictly valid for homogeneous turbulent flows, an assumption has to be made in our
case. We will assume that the turbulence is locally homogeneous so that the spatial derivatives of the turbulent statistics vanish. This is a
strong assumption, however, we are interested in a fairly good and simple approximation of the drift and diffusion matrices in order to test
stochastic models. As far as we know, there is no other simple method to determine the parameters of this particular type of stochastic models
due to the non-homogeneity of the turbulent flow studied. Moreover, it will be shown later from the stochastic simulations of the gas-solid flow
that this approximation leads to very good results. Besides, it should be also noted that it is under this assumption that \citet{walkurten}
recently derived the drift matrix of a stochastic equation predicting the fluctuating velocity of fluid particles for a turbulent pipe flow.
Nonetheless, it has to be mentioned that other stochastic models, motivated by the works of \citet{withki_BLM_2_1981}, \citet{durbin_1983}, and
\citet{thomson_84}, have been suggested to tackle the problem induced by the
non-homogeneity. More details can be found in \citet{ilihan}, \citet{ilmiha}, and references within.\\
Assuming the turbulence as locally homogeneous, the drift matrix can be expressed from \eref{eq:GLM_simonin_fluc} or \eref{eq:GLM_nous_fluc} as
\begin{equation}\label{eq:T_pope_fonc_Gij}
\mathbf{\widetilde{G}}=-\lrp{\pmb{\EuScript{T}}^T}^{-1}\,,
\end{equation}
where $(\cdot)^T$ denotes the transpose and $\EuScript{T}_{ij}$ is the matrix of the decorrelation time scales of the fluid seen which is
defined as
\begin{equation}
\EuScript{T}_{ij}=\int_{0}^{\infty} \lrc{\u_vu'_i\u_vu'_k}^{-1}\left<\u_vu'_k(0)\u_vu'_j(t)\right>\,dt\,,
\end{equation}
with $\lrc{\u_vu'_i\u_vu'_k}^{-1}$ being the $i-k$ component of the inverse of $\lrc{\mathbf{\u_vu\u_vu}^T}$. In order to obtain the drift matrix, $\EuScript{T}_{ij}$ has been computed from DNS data.\\
To determine the diffusion matrix, we have to consider the transport equation of the second order statistical moment of the fluid velocity seen by particles which is described by equation~(\ref{eq:GLM_forme_generale}). This transport equation, which can be derived from the transport equation of the joint probability density function for the particle and fluid seen velocities,\cite{minpei_01} has the following form
\begin{align}
\alpha_p\rho_p\derivtmp{}{t}\lrc{\u_vu'_i\u_vu'_j}=&-\derivp{}{x_k}\lrp{\alpha_p\rho_p\lrc{\u_vu'_i\u_vu'_jv'_{p,k}}}
-\alpha_p\rho_p\lrc{\u_vu'_iv'_{p,k}}\derivp{\lrc{u_j}}{x_k}
-\alpha_p\rho_p\lrc{\u_vu'_jv'_{p,k}}\derivp{\lrc{u_i}}{x_k}\nonumber\\
&+\alpha_p\rho_p\lrc{\u_vu'_i}\lrp{\derivp{\lrc{u'_ju'_k}}{x_k}+\frac{1}{\rho_f}\derivp{\lrc{p}}{x_j}-\nu \derivp{^2\lrc{u_j}}{x_k\partial x_k}}\nonumber\\
&+\alpha_p\rho_p\lrc{\u_vu'_j}\lrp{\derivp{\lrc{u'_iu'_k}}{x_k}+\frac{1}{\rho_f}\derivp{\lrc{p}}{x_i}-\nu \derivp{^2\lrc{u_i}}{x_k\partial x_k}}\nonumber\\
&-\alpha_p\rho_p\lrp{\lrc{v_{p,k}}-\lrc{u_{k}}}\lrp{\lrc{\u_vu'_i}\derivp{\lrc{u_j}}{x_k}+\lrc{\u_vu'_j}\derivp{\lrc{u_i}}{x_k}}\nonumber\\
&+\alpha_p\rho_p\lrc{A_i\u_vu'_j+A_j\u_vu'_i}+\alpha_p\rho_p \lrc{B_{ik}B_{jk}}\,.
 \label{eq:transp_Reyn_vu_model_1}
\end{align}
Note that this transport equation is generally written for convenience in terms of a fluctuating fluid velocity seen defined as $\u_vu''_i=\u_vu_i-\lrc{\u_vu_i}$ while we defined it as $\u_vu'_i=\u_vu_i-\lrc{u_i}$ in the present study. The other form of the transport equation can be thus found by introducing the relation $\u_vu'_i=\u_vu''_i+\lrc{\u_vu'_i}$ in \eref{eq:transp_Reyn_vu_model_1}.

Let us now write the drift vector, $A_i$, in a compact form as
 \begin{equation}\label{eq:GLM_inst_2}
A_i=-\frac{1}{\rho_f}\derivp{\lrc{p}}{x_i}+\nu \derivp{^2\lrc{u_i}}{x_j\partial
x_j}+\lrp{v_{p,j}-\u_vu_j}\derivp{\lrc{u_i}}{x_j}+G_{ij}\lrp{\u_vu_j-\lrc{u_j}}+C_i\,.
\end{equation}
When $C_i=0$, the model proposed by \citet{simonin93} is recovered while the new proposed model is obtained if
$C_i=\partial\lrp{\lrc{\u_vu'_iv'_{p,k}}-\lrc{u'_iu'_k}}/\partial x_k$. Introducing the expression of the drift vector in
\eref{eq:transp_Reyn_vu_model_1} yields
\begin{align}
\alpha_p\rho_p\derivtmp{}{t}\lrc{\u_vu'_i\u_vu'_j}=&-\derivp{}{x_k}\lrp{\alpha_p\rho_p\lrc{\u_vu'_i\u_vu'_jv'_{p,k}}}
+\alpha_p\rho_p\lrc{\u_vu'_i}\derivp{\lrc{u'_ju'_k}}{x_k}
+\alpha_p\rho_p\lrc{\u_vu'_j}\derivp{\lrc{u'_iu'_k}}{x_k}\nonumber\\
&-\alpha_p\rho_p\lrp{C_i\lrc{\u_vu'_j}+C_j\lrc{\u_vu'_i}}
+\alpha_p\rho_p\lrc{\widetilde{G}_{ik}\u_vu'_k\u_vu'_j+\widetilde{G}_{jk}\u_vu'_k\u_vu'_i}\nonumber\\
&+\alpha_p\rho_p \lrc{B_{ik}B_{jk}}\,.
 \label{eq:transp_Reyn_vu_model_2}
\end{align}
The gas-solid channel flow being statistically stationary and homogeneous in the streamwise and spanwise directions, the diffusion matrix can be expressed, under the local homogeneity assumption, from \eref{eq:transp_Reyn_vu_model_2} as a function of the drift matrix
\begin{equation}\label{eq:Gij_star_fonc_T_pope_star}
B_{ij}^2\equiv B_{ik}B_{jk}= -\widetilde{G}_{ik}\lrc{\u_vu'_k\u_vu'_j}-\widetilde{G}_{jk}\lrc{\u_vu'_k\u_vu'_i}\,.
\end{equation}
Here, it should be noted that $B_{ij}^2$ does not determine uniquely $B_{ij}$. Nevertheless, $B_{ij}^2$ will produce a unique set of statistical moments of the fluctuating fluid velocity seen by particles.\cite{thomson_87,pope_book,minpei_01} Therefore, we suppose in this study that $B_{ij}$ is symmetric.

The parameters of the Langevin model can be thus expressed as a function of the decorrelation time scales and second order statistical moment of the fluid seen by particles.
%
%
\subsection{Results}
In order to evaluate the parameters of the Langevin model, data extracted from a direct numerical simulation of a gas-solid channel flow have
been used. The direct numerical simulation was conducted at a Reynolds number $Re_b=2280$ (based on channel half-height $\delta$ and bulk
velocity $U_b$) corresponding to a Reynolds number based on the wall-shear velocity $(u_\tau)$ equals to $Re_\tau\approx 155$. The results used
in the present study are coming from the same numerical computations presented in \citet{masoku_test_case_IJMF} to test the prediction of
particle dispersion by different DNS codes. Therefore, only the main characteristics of the gas-solid flow simulation are given here. The
numerical simulation of solid particle trajectories was restricted to spherical particles smaller than the smallest turbulent length scale.
Therefore, we made use of the point-force approximation. In the present study, the particle-particle interactions as well as the turbulence
modulation were disregarded (one-way coupling). In addition, the added mass, history and lift forces were neglected in the particle equation of
motion since the ratio between the particle and fluid density obeys $\rho_p/\rho_f\gg 1$.
%
In the present study, only the non-linear drag force, estimated from the correlation of \citet{moal}, was considered.\\ Simulations were run for three sets of particles characterized by different Stokes particle response times in wall units, $\tau^+_p=1,~ 5~\text{and}~25$ [quantities in wall units are normalized with the viscous scales (i.e. the wall-shear  velocity  $u_\tau$   and  the  viscous  lengthscale $\nu/u_\tau$)  and indicated by the superscript $(\cdot)^+$]. The corresponding dimensionless diameters were $d_p/\delta=1\times10^{-3}\mathrm{,}~2.2\times10^{-3}~\mathrm{and}~5\times10^{-3}$, and the density ratio was equal to $\rho_p/\rho_f=1000/1.3$ for the three sets. Statistics on the dispersed phase were started after a time lag necessary for particle statistics to reach a stationary state.\\

\begin{figure}[!htb]
 \centering\includegraphics[width=\myfigwidth]{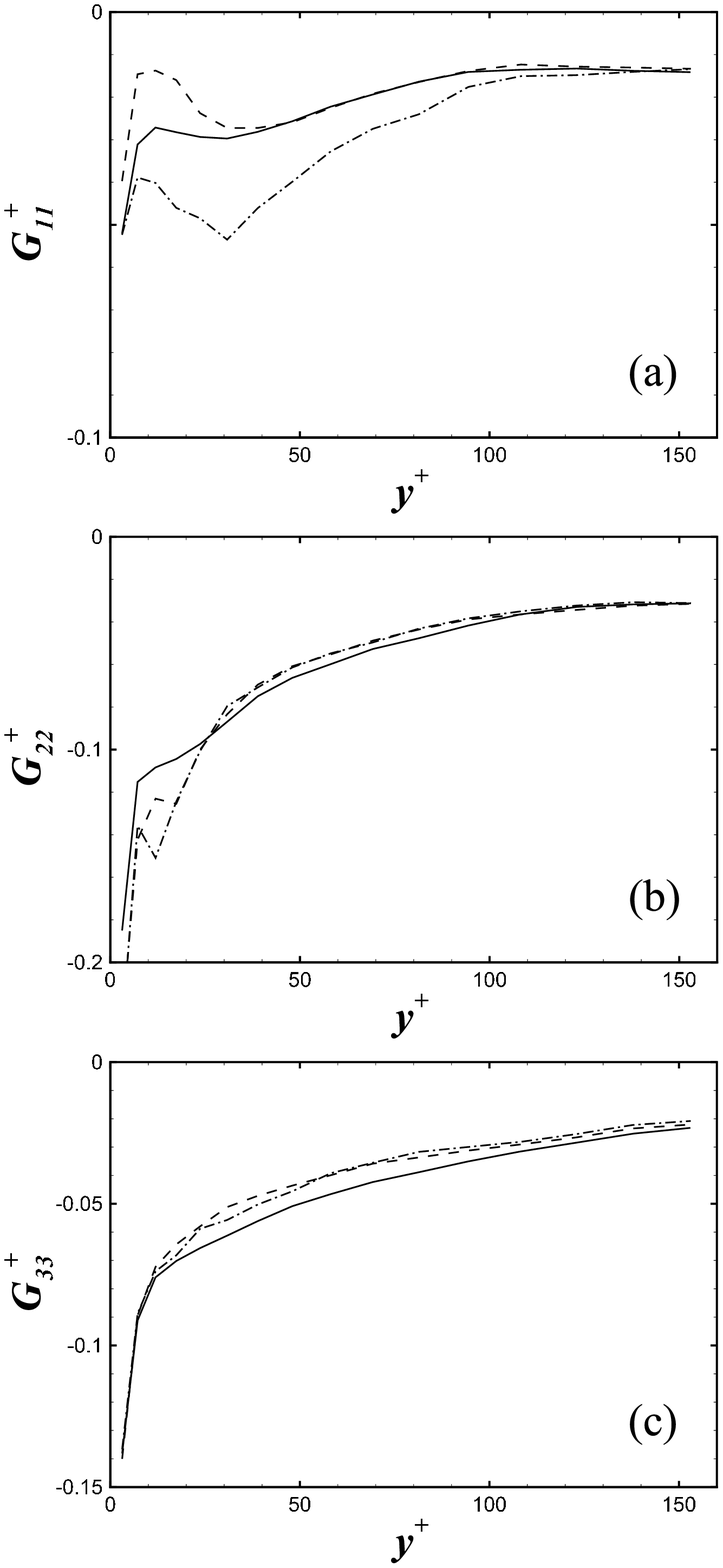}
 \caption{Diagonal components of the drift matrix, $G_{ij}$. $\tau_p^+=1$ (---) ; $\tau_p^+=5$ (--~--) ; $\tau_p^+=25$ (--~$\cdot$~--).}
\label{fig:drift_matrix_diag}
\end{figure}

\begin{figure}[!htb]
\centering\includegraphics[width=\myfigwidth]{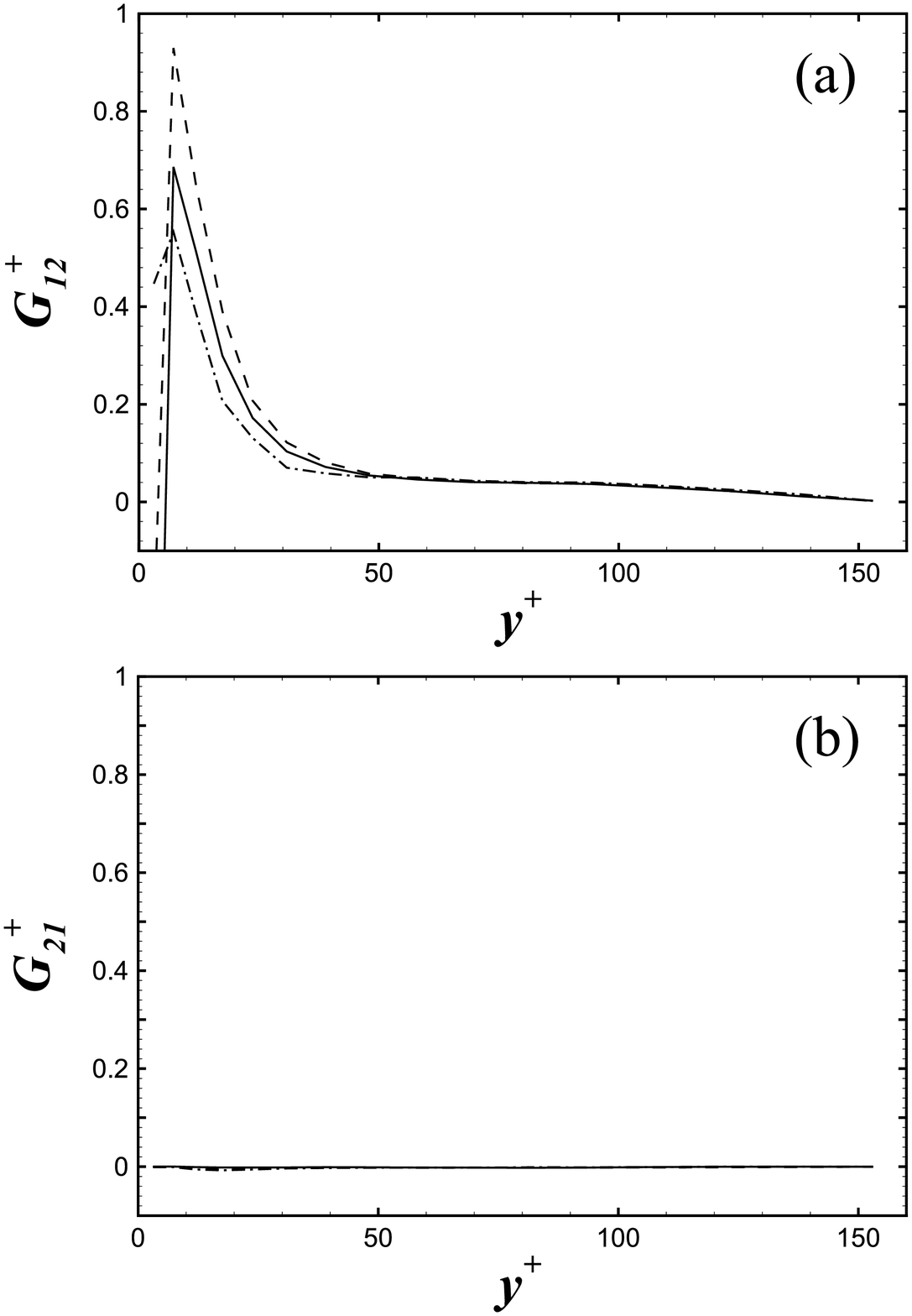}
 \caption{Non-diagonal components of the drift matrix, $G_{ij}$. $\tau_p^+=1$ (---) ; $\tau_p^+=5$ (--~--) ; $\tau_p^+=25$ (--~$\cdot$~--).}
\label{fig:drift_matrix_hors_diag}
\end{figure}

In figures~\ref{fig:drift_matrix_diag} and ~\ref{fig:drift_matrix_hors_diag}, the components of the drift matrix in wall units are plotted as a
function of the wall-normal coordinate $y^+$, and for the three different particle inertia. Before commenting the results, we have to emphasize
that not too much attention should be paid to the behavior of the Langevin model parameters near the wall since they were derived under the
assumption of local homogeneity. This approximation is certainly not correct in this region of strong gradients of the turbulent statistical
moments. From the diagonal components of $G_{ij}^{+}$ shown in Fig.~\ref{fig:drift_matrix_diag}, it can be observed that the magnitude of
$G^+_{22}$ and $G^+_{33}$ decreases monotically with increasing $y^+$. Moreover, the particle inertia is shown to not have a significant effect
on these components. These trends are quite different for $G^+_{11}$ since its magnitude is seen to have a local minimum at $y^+\approx 10$
whatever the particle inertia, and then for $y^+\apprge 40$, it decreases with increasing $y^+$. Concerning the non-diagonal components of
$G^+_{ij}$ plotted in Fig.~\ref{fig:drift_matrix_hors_diag}, we note that $G^+_{12}$ is maximum in the near-wall region and tends to zero at the
channel center. The particle inertia has a quite important effect on that component while $G^+_{21}$ is zero across the channel whatever the
particle inertia.\\ Finally, it has to be mentioned that our estimation of the drift matrix for the lowest particle inertia is qualitatively in
good agreement with the results obtained by \citet{walkurten} in their study of a Langevin model for predicting the fluctuating velocity of
fluid particles in a turbulent pipe flow.\\
%

\begin{figure}[!htb]
 \centering\includegraphics[width=.95\textwidth]{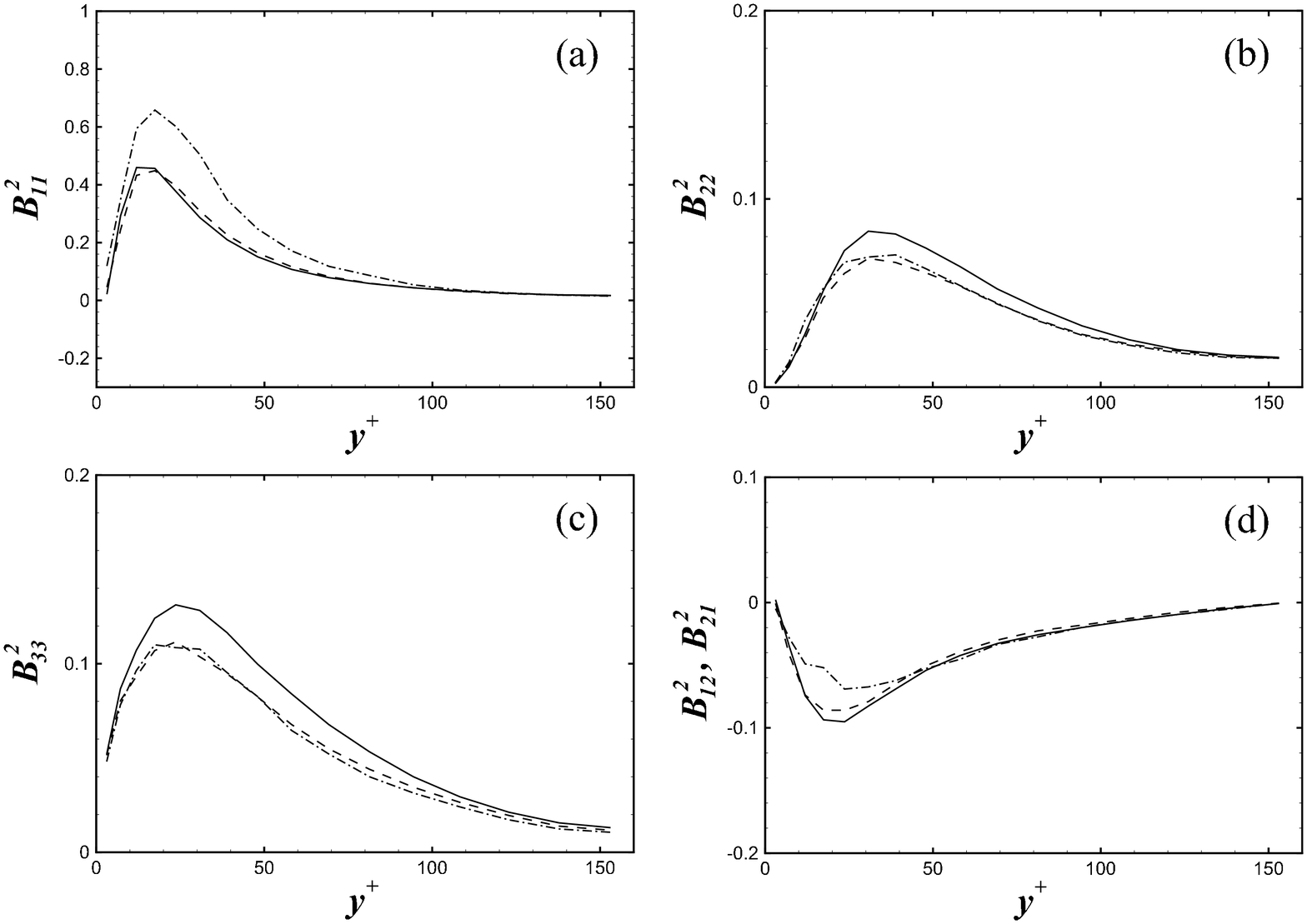}
 \caption{Components of the diffusion matrix, $B^{2}_{ij}$. $\tau_p^+=1$ (---) ; $\tau_p^+=5$ (--~--) ; $\tau_p^+=25$ (--~$\cdot$~--).}
\label{fig:diffusion_matrix}
\end{figure}

The results obtained for the diffusion matrix are given in the form of $B_{ij}^{2}=B_{ik}B_{jk}$ in figure \ref{fig:diffusion_matrix}. Contrary
to the conventional modeling assumption, it is found that the diffusion term $B^{2}_{ij}$ is significantly anisotropic for $y^+<100$. A similar
observation was previously made by \citet{pope_02} in a study of the stochastic Lagrangian modeling of fluid particle trajectories in a
homogeneous turbulent shear flow. Besides, the results show that the components of $B_{ij}^{2}$ tend towards zero close to the wall and have a
maximum located approximately at $y^+=25$. Consequently, according to these results, $B_{ij}^{2}$ cannot be modeled as a function of kinetic
energy dissipation rate moderated by a constant since the dissipation of the kinetic energy is maximum at the wall. Concerning the inertia
effect, it is observed that the values of $B_{11}^{2}$ are identical for $\tau_p^+=1\text{ and }5$ and increase when $\tau_p^+=25$. This is not
the case for the components $B_{22}^{2}$ and $B_{33}^{2}$ since similar results are obtained for $\tau_p^+=5\text{ and }25$ particles while the
magnitude of these components is higher for the lowest particle inertia. Regarding the non-diagonal component, the particle inertia effect is
seen to only change its minimum. Using these results for the drift and diffusion matrices, the performance of the present stochastic model is
examined in the next section.
%
%
\section{Evaluation of the stochastic models}
\subsection{Presentation of the test}
To assess the performance of the proposed Langevin model, a comparison between results obtained from a stochastic simulation and those extracted
from the direct numerical simulation of a gas-solid channel flow has been conducted. The stochastic simulations have been carried out for three
different forms of the drift vector in order to investigate the effects of this term on the predicted dispersed phase statistics. The
expressions of the stochastic differential equation corresponding to these models can be put in the following compact form:
 \begin{equation}\label{eq:GLM_fluc_general}
d\u_vu'_i=\widetilde{G}_{ij}\u_vu'_jdt + B_{ij} dW_j + D_idt\,.
\end{equation}
Consequently, $D_i=\partial\lrc{u'_iu'_k}/\partial x_k$ gives the SDE obtained using the model proposed by \citet{simonin93}
[\eref{eq:GLM_simonin_fluc}], $D_i=\partial\lrc{\u_vu'_iv'_{p,k}}/\partial x_k$ is the second form which is derived using the proposed model for
the drift vector [\eref{eq:GLM_nous_fluc}], and $D_i=0$ corresponds to the third model considered. This last form is less cumbersome than the
two others and does not need to know beforehand the fluid Reynolds stresses or the fluid-particle covariances. In addition, it should be noted
that this model only gives the proper limit of the average of the time variation of the fluid velocity seen when $\tau_p\gg 1$ [see
\eref{eq:transp_derive_exacte_grand_taup}].
%
%
\subsection{Stochastic simulation}
\begin{figure}[!htb]
 \centering\includegraphics[width=\myfigwidth]{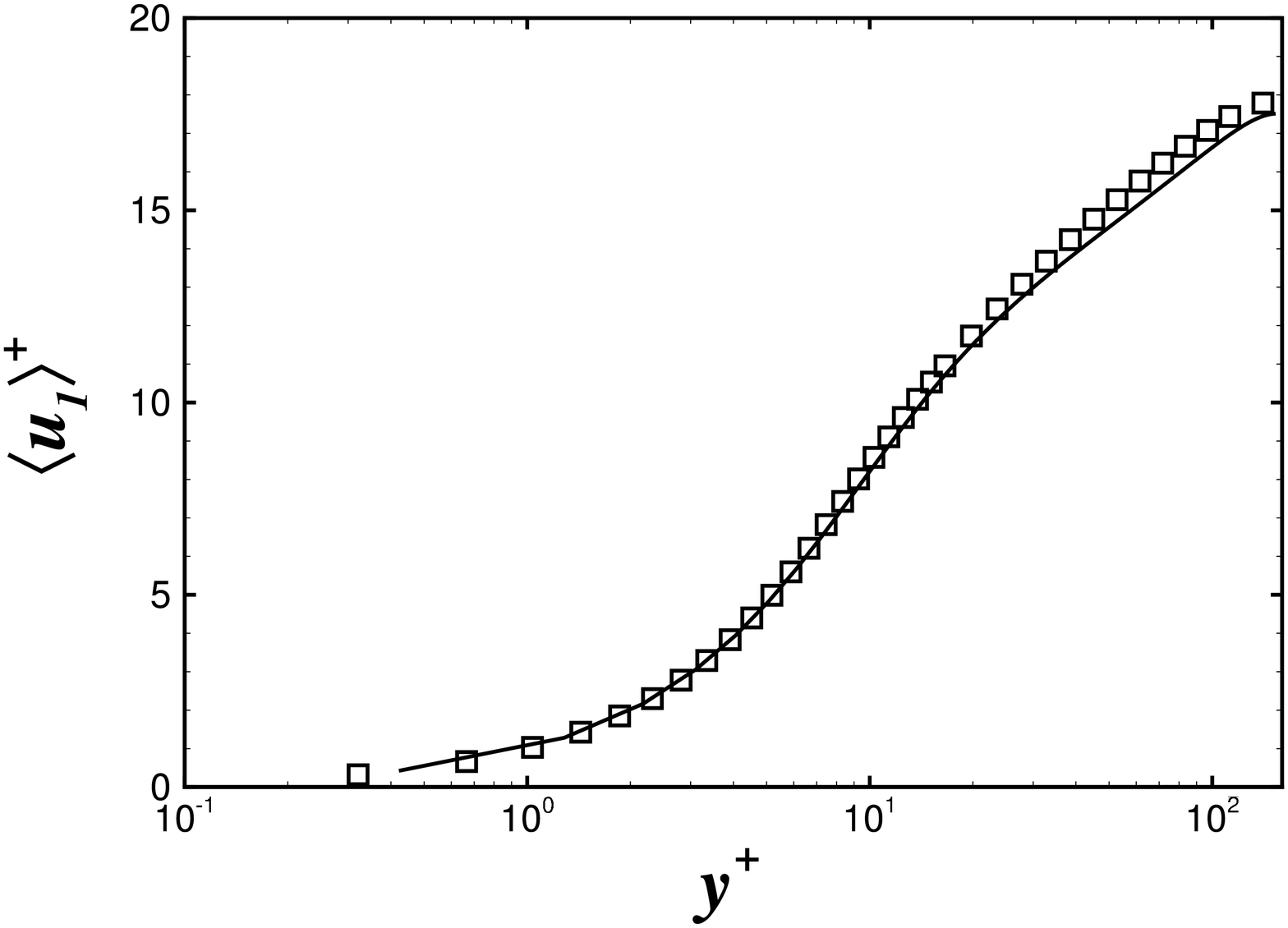}
 \caption{Mean streamwise fluid velocity, $\lrc{u_1}$. DNS (---) ; RANS ($\Box$).}
\label{fig:mean_u_RANS_DNS}
\end{figure}

For the stochastic simulation of the gas-solid flow, the mean fluid motion was calculated by means of a Reynolds-Averaged Navier-Stokes model.
Closure of the Reynolds stresses is achieved using the Non Linear Eddy Viscosity Model,\cite{spezia} so that turbulence anisotropy is taken into
account. In order to have a better precision in the near-wall region where the viscosity effects have to be taken into account, modifications of
the standard $k-\epsilon$ model following the recommendations of \citet{myokas} have been made. This model introduces ``damping'' functions that
allow the transport equations of $k$ and $\epsilon$ to be valid close to the wall. A complete description of the RANS model used here is given
in the work of \citet{cakhoe}. The mean fluid velocity, $\lrc{u_1}^+$, predicted by this model is compared in figure~\ref{fig:mean_u_RANS_DNS}
to the one obtained by DNS. Despite a slight overprediction of $\lrc{u_1}^+$ by the RANS model in the logarithmic region, the results can be
considered to be in good accordance.

After the Eulerian computation of the mean fluid velocity, the stochastic Lagrangian simulations have been performed by tracking as much as
$2.10^6$ solid particles in order to get enough statistical information in each cell of the domain to calculate the mean dispersed phase
statistics. The particle characteristics as well as the equation of motion used for calculating the trajectories are identical to those of the
DNS computation.\\
Nevertheless, contrary to the gas-solid DNS which has been conducted in a bi-periodic domain, a finite streamwise length channel has been
considered for the stochastic simulation. This length has been chosen to be equal to 10 m $(\simeq 500\text{ times the channel half-width})$ in
order to obtain dispersed statistics, calculated at the outlet, which are independent of the distance to the inlet.\\
The fluid velocity fluctuation at the particle location has been determined by integrating in time the stochastic equation
[\eref{eq:GLM_fluc_general}] in a semi-analytical way. The stochastic part is firstly disregarded and the time increment of the velocity can be
thus analytically obtained from the resulting system of coupled equations. The stochastic term increment is then estimated using an Euler scheme
and added to the analytical solution. The simulations have been carried out using a time step being equal to $\tau_p/25$ for $\tau_p^+=5\text{
and }25$. For the $\tau_p^+=1$ particles, the time step was chosen to be $\tau_p/5$ in order to limit the computational cost. One should note
that during these stochastic simulations, the time step is always lower or of the order of the smallest velocity timescale characterizing the
present flow. This choice is also in accordance with the guideline given by \citet{sommer_IJMF_2003}. The values of $\widetilde{G}_{ij}$ and
$B_{ij}$ have been linearly interpolated at the solid particle location from the data presented in the previous section. These coefficients
being unknown at the wall, a linear extrapolation has been chosen to estimate them near the walls ($0<y^+<3.1$). In addition, the mean turbulent
statistics, which appear in the three tested stochastic models, have been also calculated at the particle position using a linear interpolation
of data extracted from the DNS computation in order to not introduce additional modeling uncertainties.
%

\subsection{Numerical results}
\begin{figure}[!htb]
 \centering\includegraphics[width=\myfigwidth]{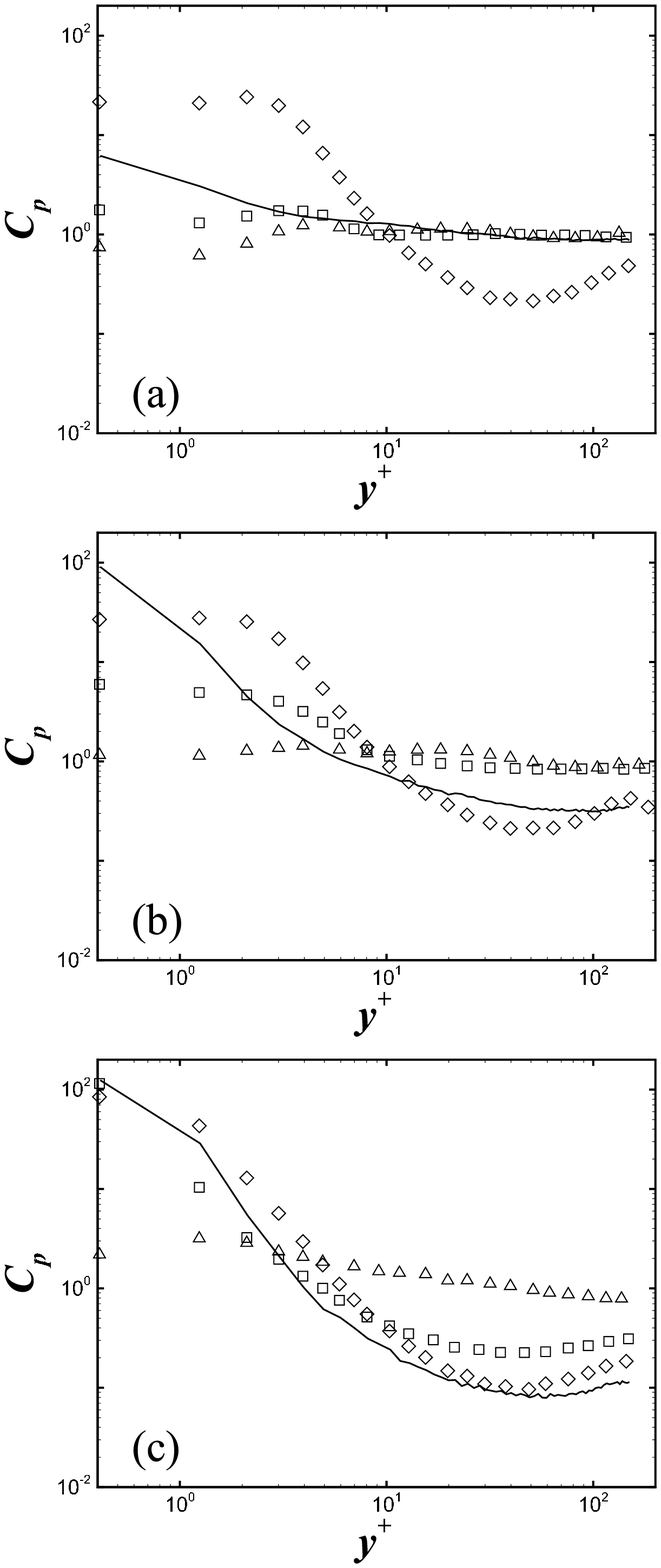}
 \caption{Particle concentration, $C_p$, for $\tau_p^+=1$ (a), $\tau_p^+=5$ (b), and $\tau_p^+=25$ (c). DNS: --- . Stochastic simulation with
 $D_i=\partial\lrc{\u_vu'_iv'_{p,k}}/\partial x_k$ ($\Box$) ;
 $D_i=\partial\lrc{u'_iu'_k}/\partial x_k$ ($\triangle$) ; $D_i=0$ ($\Diamond$).}
\label{fig:concentration_part}
\end{figure}

The first result we present is the concentration (more precisely the number density) of solid particles across the channel width. The DNS and
stochastic simulations conducted with the three different models are compared in figures~\ref{fig:concentration_part}(a-c) for $\tau^+_p=1,~
5~\text{and}~25$ respectively. These results are interesting since they reveal that the drift model has an important effect on the particle
distribution in the channel. The DNS data show that the particle concentration increases with increasing inertia (in the particle inertia range
studied). This behavior can be seen as a preferential concentration effect at the macroscopic scale. Of course, it is different from the the
local effect which occurs at smaller length scales.\cite{squires-eaton-91,wang_maxey,rouson,pimaso}\\
The comparison with the stochastic simulation shows also that the model of the drift vector proposed by \citet{simonin93}
[\eref{eq:GLM_fluc_general} with $D_i=\partial\lrc{u'_iu'_k}/\partial x_k$] is not able to reproduce the accumulation of the larger particles in
the low turbulent intensity regions. The concentration profiles remain quite uniform whatever the particle inertia. It seems that the presence
in the SDE of the term $D_i=\partial\lrc{u'_iu'_k}/\partial x_k$ (which does not vary as a function of particle inertia) prevents the accumulation.
On contrary, the model with $D_i=0$ produces a segregation of the particles in the near-wall region even for the lowest particle inertia. This
is in contradiction with the law of conservation of mass since these particles, which can be assimilated to fluid particle tracers, have to
approximately be uniformly distributed.\cite{thomson_87} This non-physical behavior, which is called spurious drift effect, was observed by
\citet{withki_BLM_2_1981} and \citet{MacBra_POF_1992} in stochastic Lagrangian simulations of tracer particles in non-homogeneous turbulent
flow. In fact, it is due to an inconsistency between the stochastic Lagrangian model and the Navier-Stokes equations which causes a
misrepresentation of the averaged time derivative of the fluctuating fluid particle velocity. A more detailed presentation of this effect can be
found in \citet{pope_87}, \citet{thomson_87}, \citet{gui_min_pof_2008}, and references within. Despite this major problem for low particle
inertia, this stochastic model predicts reasonably well the concentration of the $\tau_p^+=25$ particles. This confirms that this latter model
should be more appropriate to estimate the fluid velocity seen by large particle inertia. Finally, it is noted that the results obtained with
the present model for $\tau_p^+=1\text{ and }25$ are in good accordance with the DNS data while important discrepancies are observed for
$\tau_p^+=5$ when $y^+<2$. Nevertheless, the main point is that this model reproduces qualitatively quite well the effect of inertia on the
particle concentration. There is no spurious drift effect for low inertia and an increase of the concentration in the near-wall region is noted
for the higher particle inertia. This result is of importance since the correct prediction of particle flux in wall-bounded turbulent flows is
decisive when studying the complex process of particle deposition.\\

\begin{figure}[!htb]
 \centering\includegraphics[width=\myfigwidth]{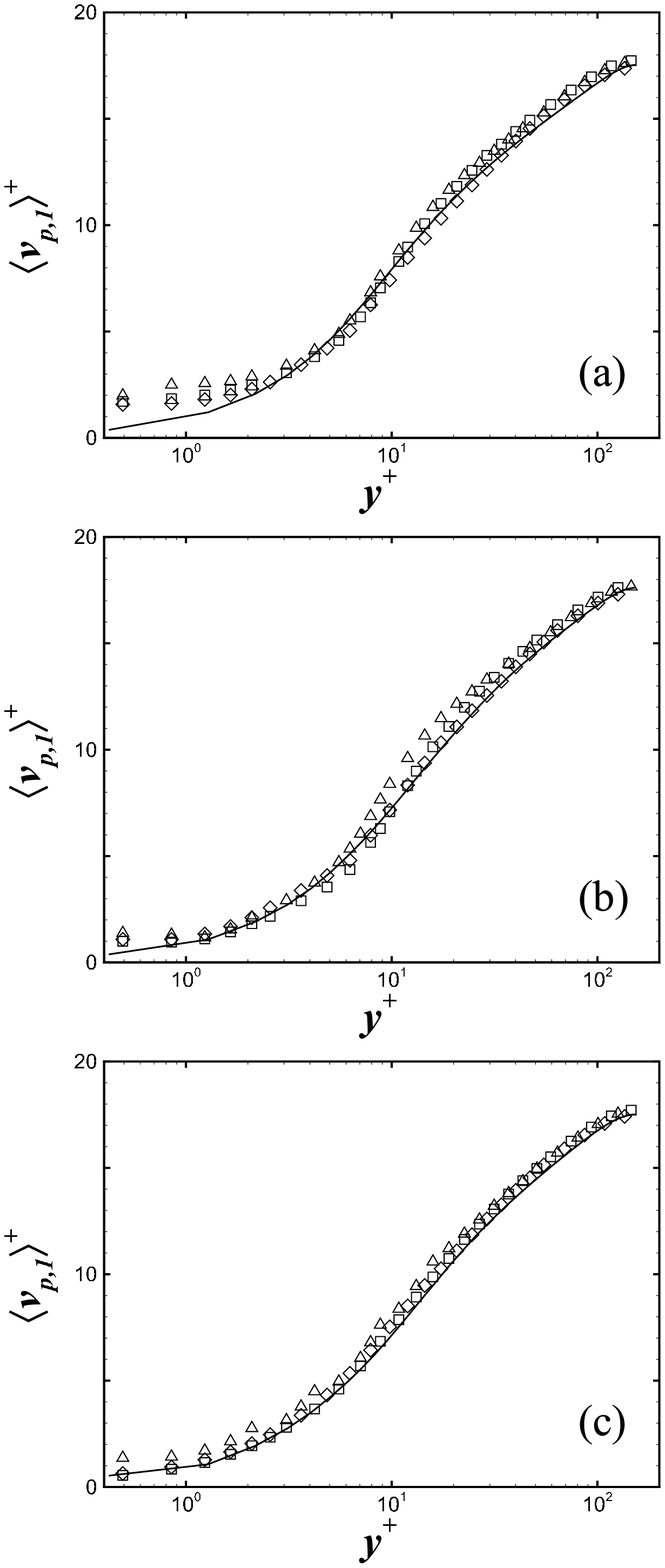}
 \caption{Mean streamwise particle velocity, $\lrc{v_{p,1}}$, for $\tau_p^+=1$ (a), $\tau_p^+=5$ (b), and $\tau_p^+=25$ (c). DNS: --- .
 Stochastic simulation with $D_i=\partial\lrc{\u_vu'_iv'_{p,k}}/\partial x_k$ ($\Box$) ;
 $D_i=\partial\lrc{u'_iu'_k}/\partial x_k$ ($\triangle$) ; $D_i=0$ ($\Diamond$).}
\label{fig:vitesse_moyenne_part}
\end{figure}

The first order statistical moment of the particle velocity is plotted in figures~\ref{fig:vitesse_moyenne_part}(a-c). From
Fig.~\ref{fig:vitesse_moyenne_part}(a), it is seen that the mean velocity of the smallest particle inertia studied is quite well predicted
across the channel by the different models of the drift vector. Nevertheless, this velocity is overestimated in the viscous sublayer for
$y^+<2$. The observed discrepancy can be reasonably attributed to the combination of two approximations. The first one concerns the local
homogeneity assumption made to derive the parameters of the stochastic which does not hold in this region. The second one is the linear
extrapolation used to estimate these parameters at the particle location. These remarks should be kept in mind throughout the presentation of
the results. In the buffer and logarithmic regions, the present model as well as the one of \citet{simonin93} give similar results which are
slightly greater than those of the DNS. This is due to the fact that the RANS model slightly overestimates the mean fluid velocity given by the
DNS. Concerning the results obtained with the third model tested, i.e. \eref{eq:GLM_fluc_general} with $D_i=0$, we note that they are similar to
those of the two other models except in the buffer region where the mean particle is lower. It is believed that this interesting difference is
due to the fact that this model is not compatible with the transport equation of the drift velocity in the limit of low particle inertia. This
incompatibility certainly gives rise to a wrong estimation of the drift velocity which should be quite low for this kind of particles. Since the
mean particle velocity is lower than expected, it can concluded that this model generates fluctuations of the fluid velocity whose average is
negative.\\
The mean velocity of the $\tau_p^+=5$ particles is shown in Fig.~\ref{fig:vitesse_moyenne_part}(b). Firstly, the present model and the one with
$D_i=0$ give identical results. The mean particle velocity is also slightly overestimated in the buffer and logarithmic regions. Nonetheless,
the discrepancy observed for the smaller particles in the near-wall region is attenuated. This is certainly due to the less important
sensitivity of the $\tau_p^+=5$ particles to the fluctuating fluid velocity. Secondly, it is noticed that the model of \citet{simonin93} causes
a too high particle velocity in the buffer region. The incompatibility of this model with the transport equation of the drift velocity in the
limit of large particle inertia can be invoked. Nevertheless, this explanation has to be taken with caution since one could argue that
$\tau_p^+=5$ particles do not belong to the category of large particles. At this stage of the study, no clear conclusion can be drawn.\\
The mean velocity of $\tau_p^+=25$ particles is plotted in Fig.~\ref{fig:vitesse_moyenne_part}(c), the three models give approximately the
same results in the buffer and logarithmic regions. The predicted velocity is once more slightly higher than the one extracted from the DNS. In
the viscous sublayer, this overprediction is also observed for the model of \citet{simonin93} while the results with
the two other models are in a good accordance with the DNS.\\

\begin{figure}[!htb]
 \centering\includegraphics[width=\myfigwidth]{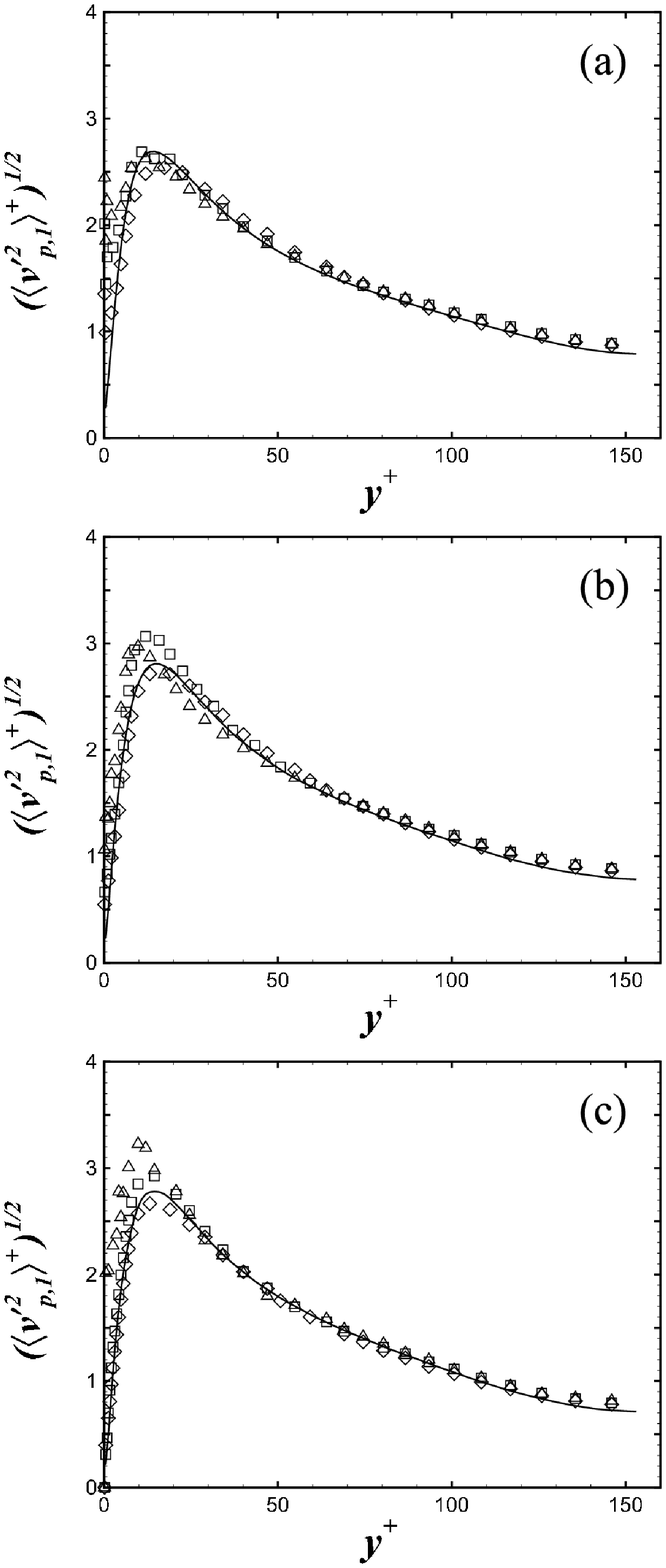}
 \caption{Root mean square of the streamwise particle velocity, $(\lrc{v'^2_{p,1}})^{1/2}$, for $\tau_p^+=1$ (a), $\tau_p^+=5$ (b), and $\tau_p^+=25$ (c). DNS: --- . Stochastic simulation with
 $D_i=\partial\lrc{\u_vu'_iv'_{p,k}}/\partial x_k$ ($\Box$) ;
 $D_i=\partial\lrc{u'_iu'_k}/\partial x_k$ ($\triangle$) ; $D_i=0$ ($\Diamond$).}
\label{fig:vitesse_rms_part_1}
\end{figure}

\begin{figure}[!htb]
 \centering\includegraphics[width=\myfigwidth]{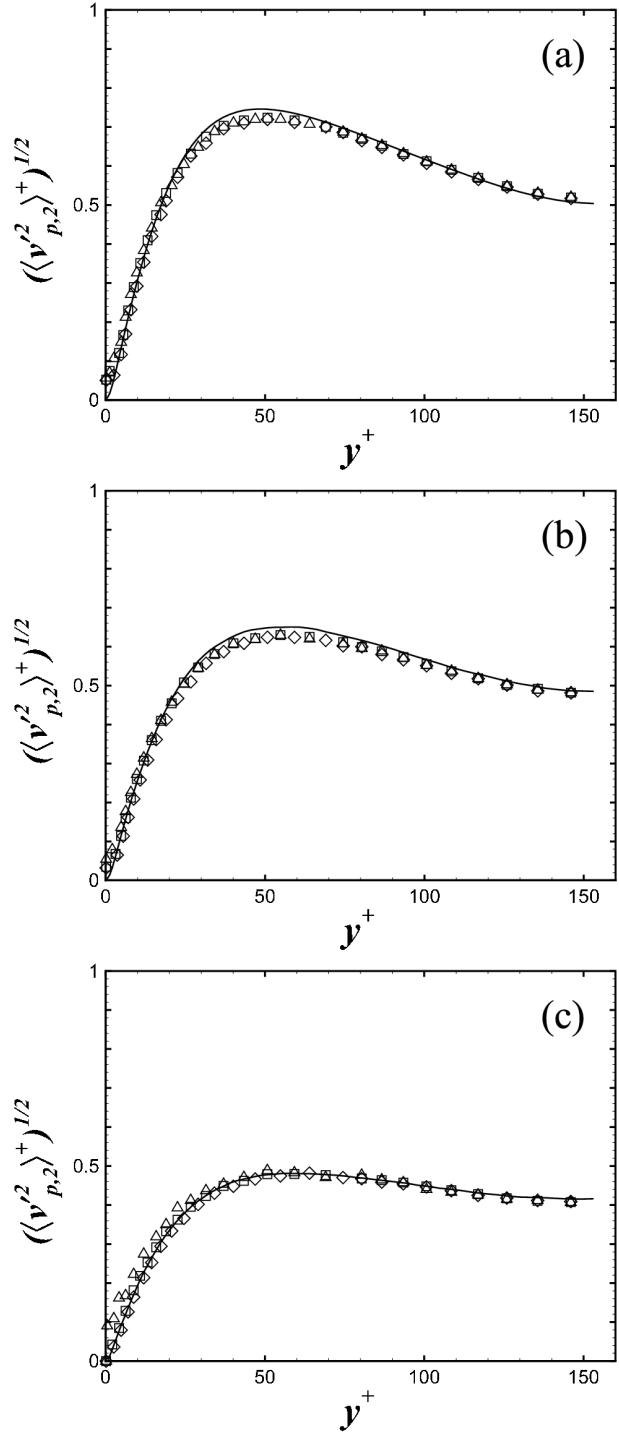}
 \caption{Root mean square of the wall-normal particle velocity, $(\lrc{v'^2_{p,2}})^{1/2}$, for $\tau_p^+=1$ (a), $\tau_p^+=5$ (b), and $\tau_p^+=25$ (c). DNS: --- . Stochastic simulation with
 $D_i=\partial\lrc{\u_vu'_iv'_{p,k}}/\partial x_k$ ($\Box$) ;
 $D_i=\partial\lrc{u'_iu'_k}/\partial x_k$ ($\triangle$) ; $D_i=0$ ($\Diamond$).}
\label{fig:vitesse_rms_part_2}
\end{figure}

\begin{figure}[!htb]
 \centering\includegraphics[width=\myfigwidth]{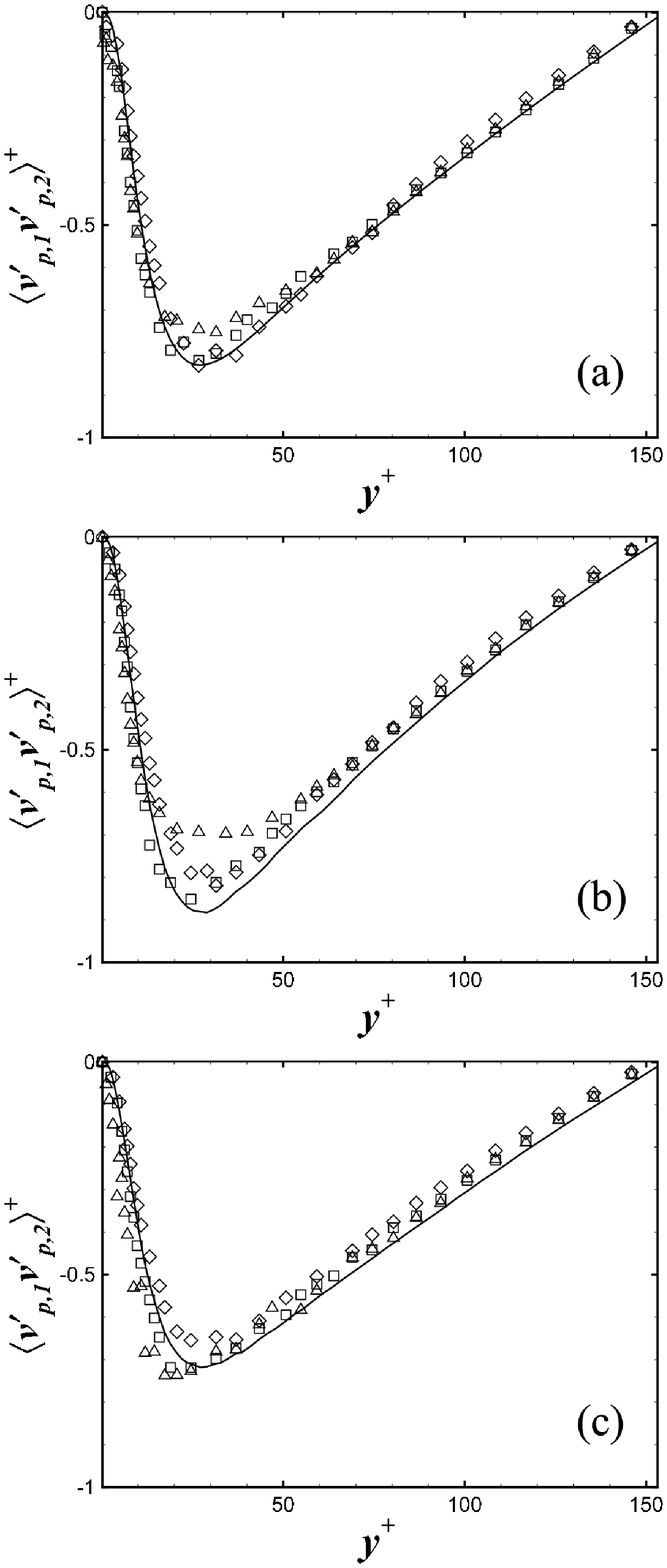}
 \caption{Particle kinetic shear stress, $\lrc{v'_{p,1}v'_{p,2}}$, for $\tau_p^+=1$ (a), $\tau_p^+=5$ (b), and $\tau_p^+=25$ (c). DNS: --- .
 Stochastic simulation with $D_i=\partial\lrc{\u_vu'_iv'_{p,k}}/\partial x_k$ ($\Box$) ; $D_i=\partial\lrc{u'_iu'_k}/\partial x_k$ ($\triangle$)
 ; $D_i=0$ ($\Diamond$).}
\label{fig:vitesse_rms_part_12}
\end{figure}

In order to better understand the influence of this three different stochastic models, the particle velocity root mean square (rms) has also
been computed. The results obtained are shown in figures~\ref{fig:vitesse_rms_part_1}-\ref{fig:vitesse_rms_part_12}. The estimated values of the
streamwise particle velocity rms for the $\tau_p^+=1$ particles [Fig.~\ref{fig:vitesse_rms_part_1}(a)] are seen to be in very good accordance
with the DNS results. For the $\tau_p^+=5$ particles [Fig.~\ref{fig:vitesse_rms_part_1}(b)], some differences are noted for $y^+<30$.
Surprisingly, better results are obtained with the simplest model studied [i.e. \eref{eq:GLM_fluc_general} with $D_i=0$]. The two other models
overestimate the streamwise particle velocity rms. The difference is roughly of the order of 5\%. The results obtained for $\tau_p^+=25$
[Fig.~\ref{fig:vitesse_rms_part_1}(c)] show that the present model and the one with $D_i=0$ predict accurately this particle velocity
statistical moment. Some important discrepancies are noticed with the model by \citet{simonin93} for $y^+<15$. Concerning the wall-normal
component plotted in figure~\ref{fig:vitesse_rms_part_2}, it can be seen that the accordance with the DNS results is very good whatever the
model used. The predicted values are slightly lower than those given by the DNS for $\tau_p^+=1\text{ and } 5$, however, this difference is not
significant. The inertia filtering effect is consequently well reproduced by the stochastic models. For sake of conciseness and due to its
strong qualitative similarity with the wall-normal component, the spanwise velocity rms is not shown. The non-diagonal component of the particle
velocity rms is presented in Fig.~\ref{fig:vitesse_rms_part_12}. More important differences are observed for this component than for the others.
Concerning the smallest particle inertia, the three models are in good agreement with the DNS data. For higher inertia, the results obtained
from these models diverge for $10<y^+<50$. In this region, the model by \citet{simonin93} underestimates the magnitude of the minimum of
particle kinetic shear stress $\lrc{v'_{p,1}v'_{p,2}}$. This trend is also obtained with the two other models, however, the difference is less.
It should be also noted that the magnitude of $\lrc{v'_{p,1}v'_{p,2}}$ given by DNS is higher than that of the three stochastic models for
$y^+>50$. A possible reason for this disagreement is that the wall-shear velocity used to normalized the quantities shown is not perfectly
identical in the DNS and stochastic simulations. Finally, it can be remarked that the differences noticed previously are less for the
$\tau_p^+=25$ particle except for $y^+>50$ where the underestimation of the magnitude
particle kinetic shear stress is of the same order.\\

\begin{figure}[!htb]
 \centering\includegraphics[width=\myfigwidth]{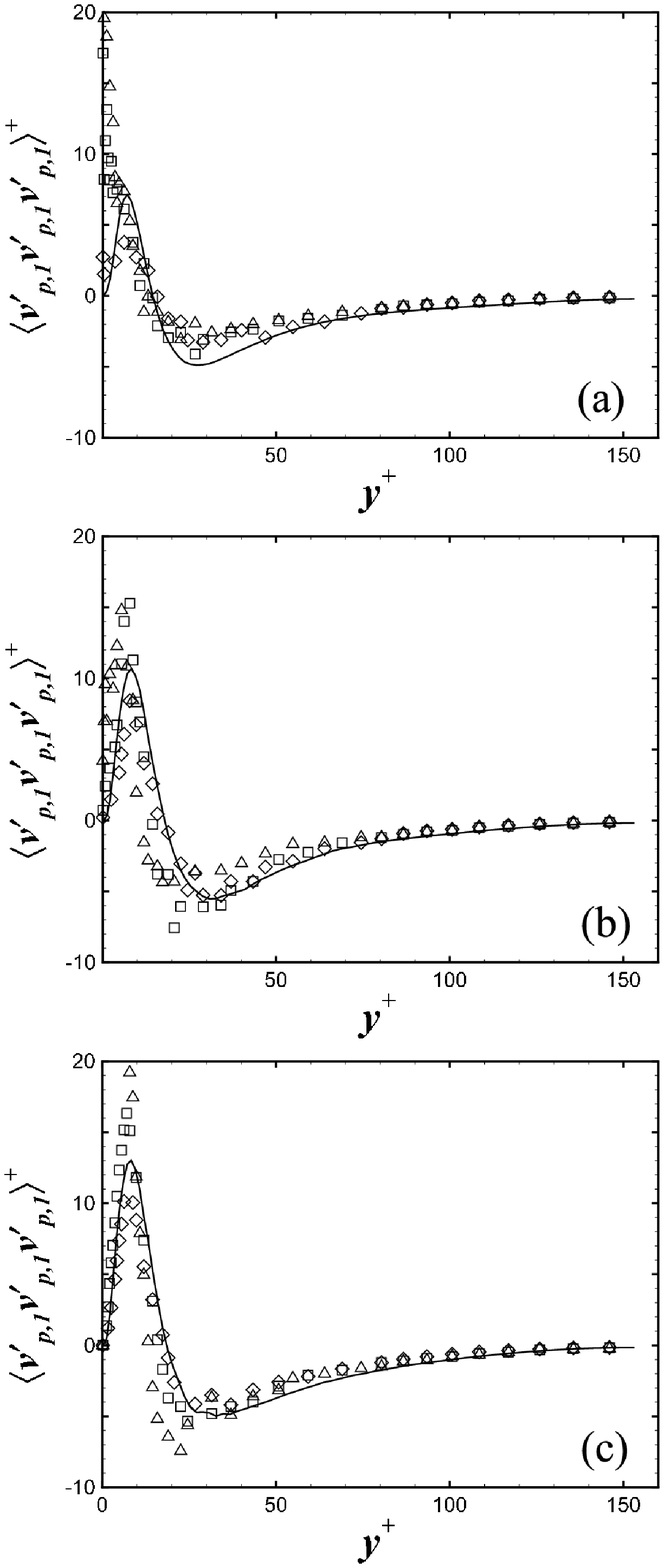}
 \caption{Triple particle velocity correlation, $\lrc{v'_{p,1}v'_{p,1}v'_{p,1}}$, for $\tau_p^+=1$ (a), $\tau_p^+=5$ (b), and $\tau_p^+=25$ (c).
 DNS: --- . Stochastic simulation with $D_i=\partial\lrc{\u_vu'_iv'_{p,k}}/\partial x_k$ ($\Box$) ; $D_i=\partial\lrc{u'_iu'_k}/\partial x_k$
 ($\triangle$) ; $D_i=0$ ($\Diamond$).}
\label{fig:vitesse_triple_part_111}
\end{figure}

\begin{figure}[!htb]
 \centering\includegraphics[width=\myfigwidth]{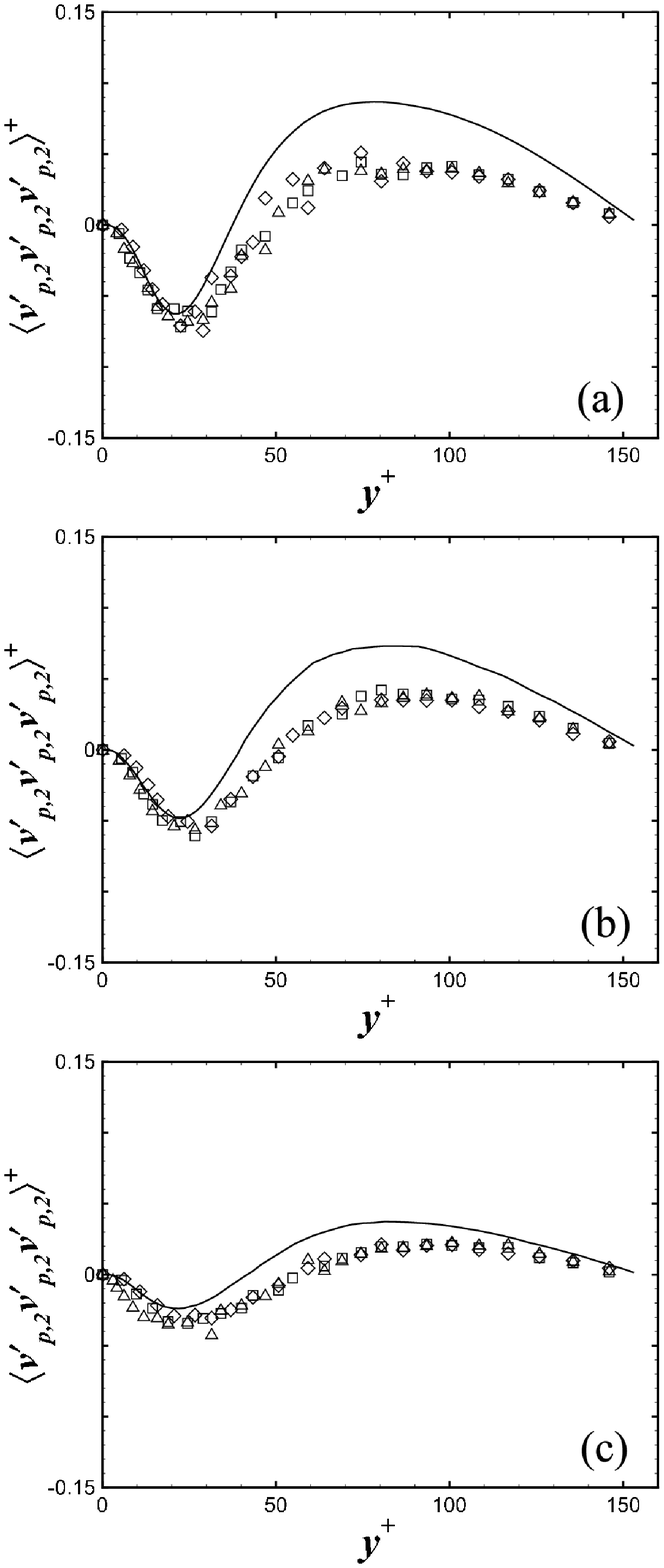}
 \caption{Triple particle velocity correlation, $\lrc{v'_{p,2}v'_{p,2}v'_{p,2}}$, for $\tau_p^+=1$ (a), $\tau_p^+=5$ (b), and $\tau_p^+=25$ (c).
 DNS: --- . Stochastic simulation with $D_i=\partial\lrc{\u_vu'_iv'_{p,k}}/\partial x_k$ ($\Box$) ; $D_i=\partial\lrc{u'_iu'_k}/\partial x_k$
 ($\triangle$) ; $D_i=0$ ($\Diamond$).}
\label{fig:vitesse_triple_part_222}
\end{figure}

\begin{figure}[!htb]
 \centering\includegraphics[width=\myfigwidth]{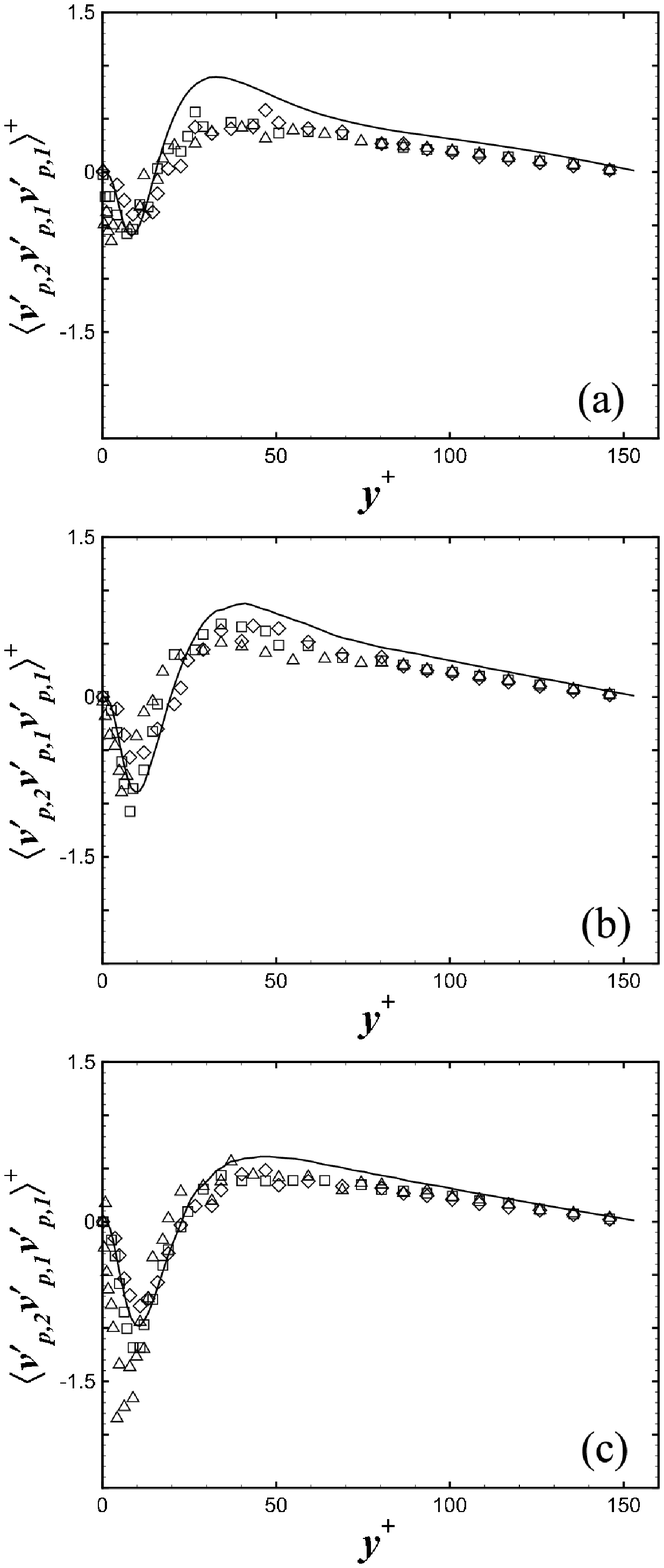}
 \caption{Triple particle velocity correlation, $\lrc{v'_{p,2}v'_{p,1}v'_{p,1}}$, for $\tau_p^+=1$ (a), $\tau_p^+=5$ (b), and $\tau_p^+=25$ (c).
 DNS: --- . Stochastic simulation with $D_i=\partial\lrc{\u_vu'_iv'_{p,k}}/\partial x_k$ ($\Box$) ; $D_i=\partial\lrc{u'_iu'_k}/\partial x_k$
 ($\triangle$) ; $D_i=0$ ($\Diamond$).}
\label{fig:vitesse_triple_part_211}
\end{figure}

\begin{figure}[!htb]
 \centering\includegraphics[width=\myfigwidth]{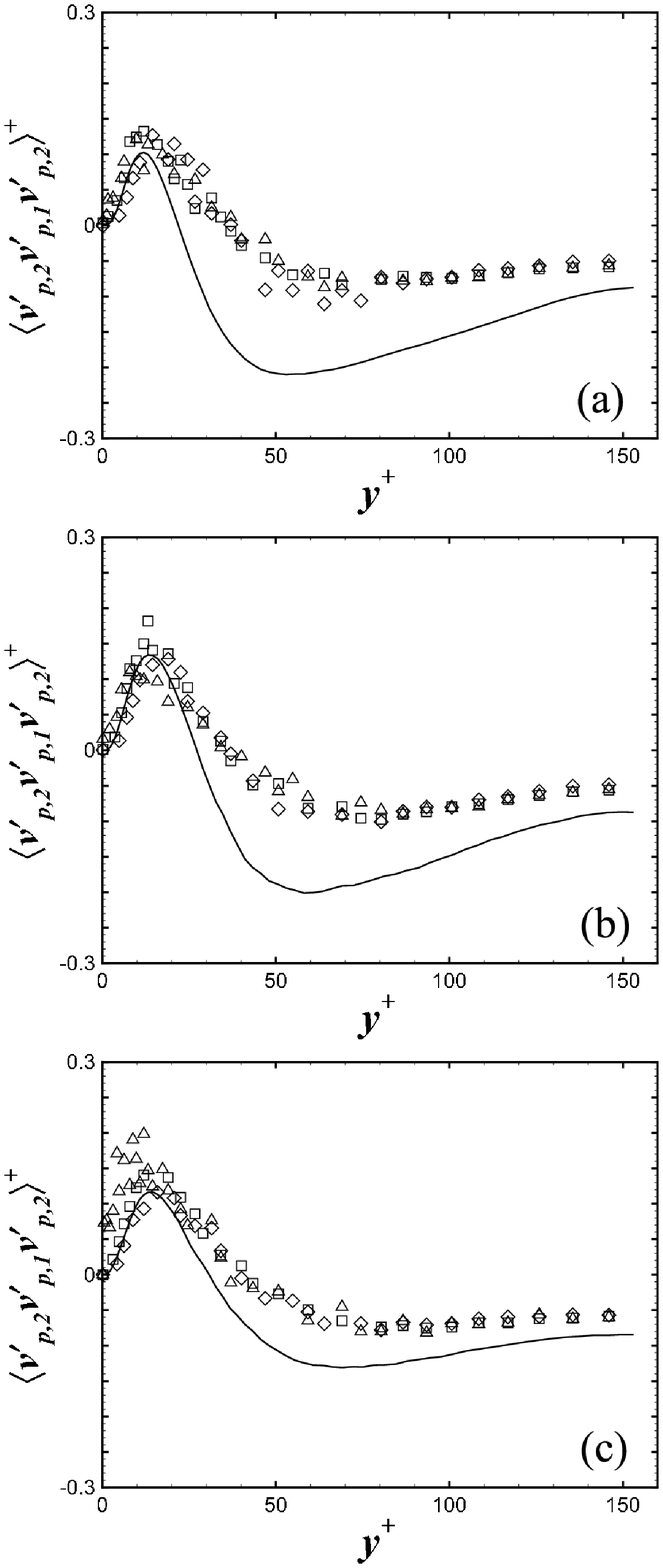}
 \caption{Triple particle velocity correlation, $\lrc{v'_{p,2}v'_{p,1}v'_{p,2}}$, for $\tau_p^+=1$ (a), $\tau_p^+=5$ (b), and $\tau_p^+=25$ (c).
 DNS: --- . Stochastic simulation with $D_i=\partial\lrc{\u_vu'_iv'_{p,k}}/\partial x_k$ ($\Box$) ; $D_i=\partial\lrc{u'_iu'_k}/\partial x_k$
 ($\triangle$) ; $D_i=0$ ($\Diamond$).}
\label{fig:vitesse_triple_part_212}
\end{figure}

\begin{figure}[!htb]
 \centering\includegraphics[width=\myfigwidth]{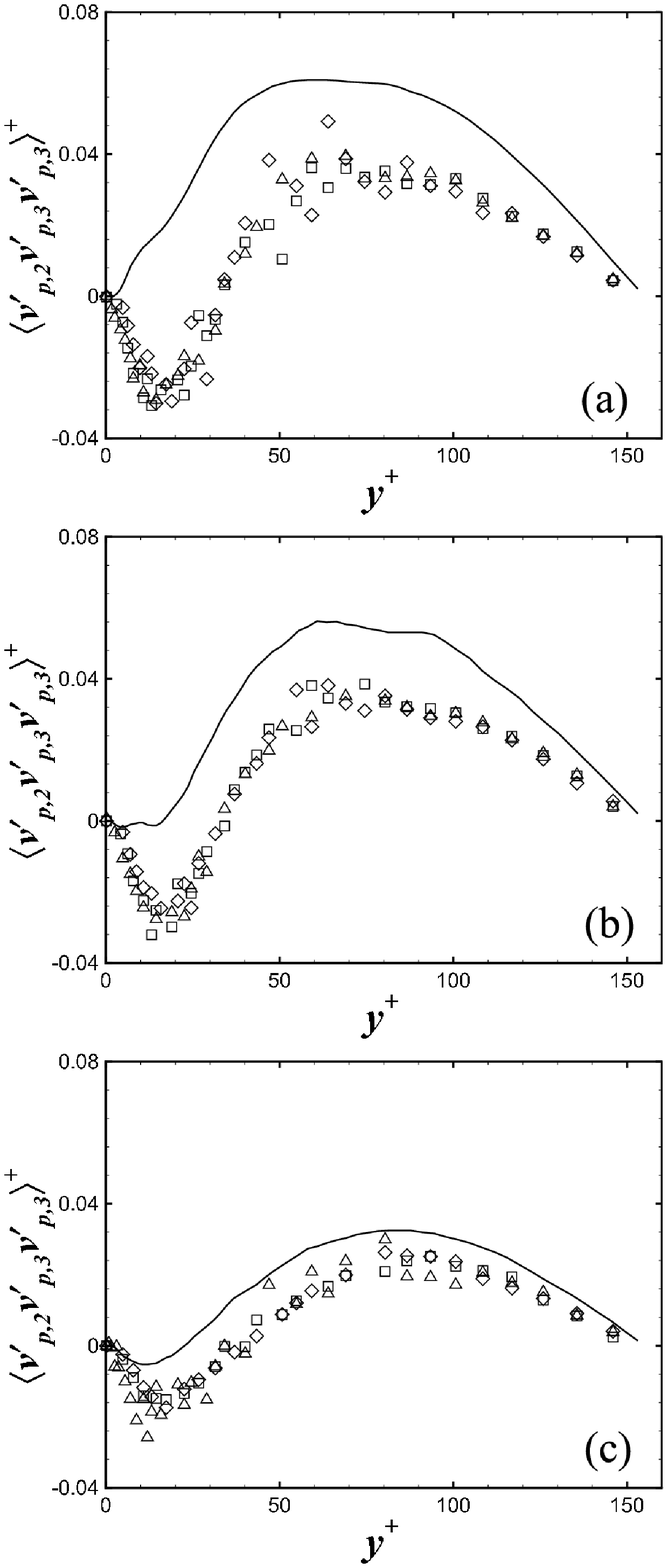}
 \caption{Triple particle velocity correlation, $\lrc{v'_{p,2}v'_{p,3}v'_{p,3}}$, for $\tau_p^+=1$ (a), $\tau_p^+=5$ (b), and $\tau_p^+=25$ (c).
 DNS: --- . Stochastic simulation with $D_i=\partial\lrc{\u_vu'_iv'_{p,k}}/\partial x_k$ ($\Box$) ; $D_i=\partial\lrc{u'_iu'_k}/\partial x_k$
 ($\triangle$) ; $D_i=0$ ($\Diamond$).}
\label{fig:vitesse_triple_part_233}
\end{figure}

To complete this comparison of the particle velocity statistics, the prediction of a higher statical moment (the triple particle velocity
correlations $\lrc{v'_{p,i}v'_{p,j}v'_{p,k}}$) has been also investigated. This will give an idea of the capability of stochastic modeling. The
agreement with the DNS data is not expected to be as good as for the other statistical moments shown due to the assumptions made for the
derivation of the parameters of the stochastic models. The streamwise triple particle velocity correlation is plotted in
figure~\ref{fig:vitesse_triple_part_111}. Surprisingly, the three stochastic models are seen to be able to predict very well this correlation
for $y^+>30$ and whatever the particle inertia. Moreover, the models well estimate the inertia effect since the obtained maximum of this
correlation (located at $y^+\simeq 10$) increases with increasing inertia. Nonetheless, the magnitude of this maximum is not well predicted
since the present model as well as the one by \citet{simonin93} clearly overestimate it while an underestimation is noted for the model with
$D_i=0$. From figure~\ref{fig:vitesse_triple_part_211}, it can be observed that $\lrc{v'_{p,2}v'_{p,1}v'_{p,1}}$ is also quite well predicted by
the present model and the one with $D_i=0$ across the channel and whatever the particle inertia. Concerning the model of \citet{simonin93}, more
important discrepancies arise for $y^+<20$ and $\tau_p^+=5\text{ and } 25$. For the correlations $\lrc{v'_{p,2}v'_{p,2}v'_{p,2}}$ and
$\lrc{v'_{p,2}v'_{p,1}v'_{p,2}}$, shown in figures~\ref{fig:vitesse_triple_part_222} and \ref{fig:vitesse_triple_part_212}, a similar trend is
noted. First, the results given by the three stochastic models are almost identical. Secondly, the agreement with the DNS data is very good when
$y^+<20$ whereas the magnitude of these two correlations is significantly underestimated in the rest of the channel. The last non-zero
correlation computed is $\lrc{v'_{p,2}v'_{p,3}v'_{p,3}}$ (Fig.~\ref{fig:vitesse_triple_part_233}). There are major differences between the DNS
and stochastic simulations. The predicted correlation is generally smaller than the one extracted from the DNS. In addition, for $y^+<40$ and
the smallest particle inertia, the wrong sign of $\lrc{v'_{p,2}v'_{p,3}v'_{p,3}}$ is given by the stochastic models. Nonetheless, the results
obtained with these models are in a good qualitative agreement for $y^+>40$, and the inertia effect, which causes a decrease of this
correlation, is well estimated.\\

\begin{figure}[!htb]
 \centering\includegraphics[width=\myfigwidth]{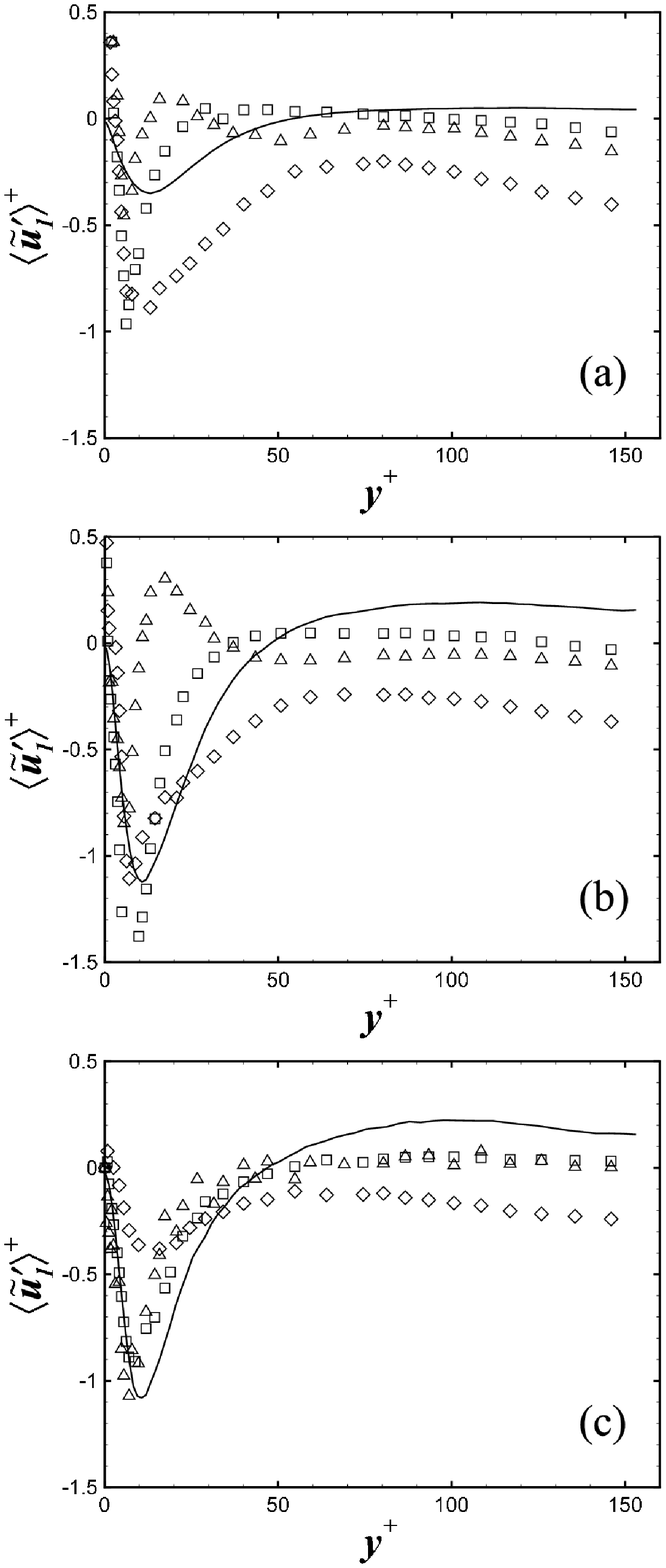}
 \caption{Streamwise drift velocity, $\lrc{\u_vu'_1}$, for $\tau_p^+=1$ (a), $\tau_p^+=5$ (b), and $\tau_p^+=25$ (c). DNS: --- . Stochastic
 simulation with $D_i=\partial\lrc{\u_vu'_iv'_{p,k}}/\partial x_k$ ($\Box$) ; $D_i=\partial\lrc{u'_iu'_k}/\partial x_k$ ($\triangle$) ; $D_i=0$
 ($\Diamond$).}
\label{fig:vitesse_derive_part_1}
\end{figure}

\begin{figure}[!htb]
 \centering\includegraphics[width=\myfigwidth]{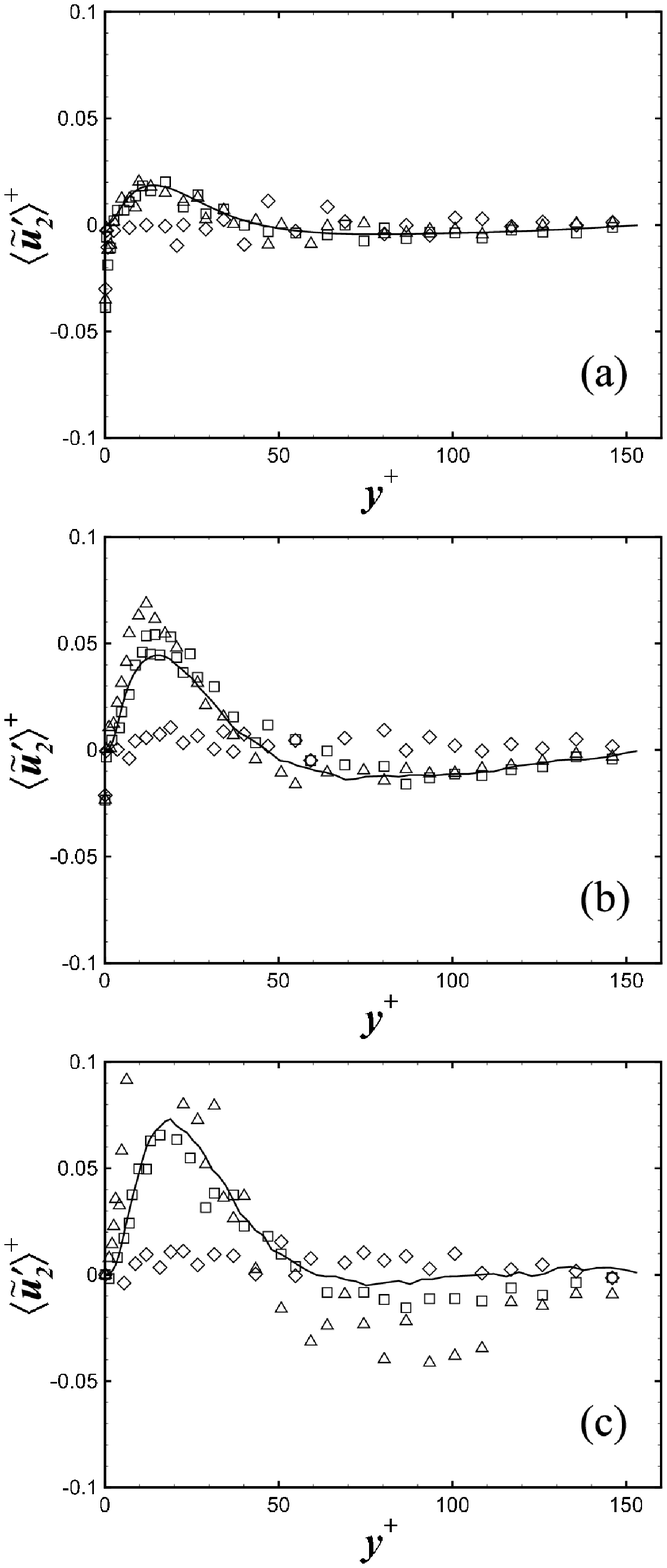}
 \caption{Wall-normal drift velocity, $\lrc{\u_vu'_2}$, for $\tau_p^+=1$ (a), $\tau_p^+=5$ (b), and $\tau_p^+=25$ (c). DNS: --- . Stochastic
 simulation with $D_i=\partial\lrc{\u_vu'_iv'_{p,k}}/\partial x_k$ ($\Box$) ; $D_i=\partial\lrc{u'_iu'_k}/\partial x_k$ ($\triangle$) ; $D_i=0$
 ($\Diamond$).}
\label{fig:vitesse_derive_part_2}
\end{figure}

The second part of this comparison between the DNS and stochastic simulations is devoted to the statistics of the fluid seen velocity. The first
statistical moment studied is the drift velocity $\lrc{\u_vu'_i}$. In figures~\ref{fig:vitesse_derive_part_1} and
\ref{fig:vitesse_derive_part_2}, the streamwise and wall-normal components are presented. A better prediction of the drift velocity from the
present model is expected since it has been previously shown that it is the only model of the three considered which is compatible with the
transport equation of this velocity in the limits of low and high particle inertia. From the results obtained for the streamwise component, it
can be seen that the three models are unable to correctly estimate it for the $\tau_p^+=1$ particles when $y^+<40$. The present model and the
one of \citet{simonin93} predict quite correctly $\lrc{\u_vu'_1}$ in the rest of the channel while significant difference with the DNS data are
still noted using the model with $D_i=0$. This is in line with the fact that this latter model is not compatible with the transport equation of
the drift velocity when $\tau_p^+\rightarrow 0$. As shown in the first part of this study, the two other models become identical in this limit.
Nevertheless, there are some differences. This is due to the value of the particle inertia studied which is not enough low to observe the
convergence of these two models. It is confirmed by the DNS data since the drift velocity of the $\tau_p^+=1$ particles is non-null while this
velocity has to vanish when $\tau_p^+\rightarrow 0$. From figure~\ref{fig:vitesse_derive_part_1}(b), it is apparent that the model with $D_i=0$
better predicts the drift velocity for this kind of particles, however, there are important inconsistencies in a large part of the channel. The
results given by the present model are in satisfactory agreement with the DNS data, whereas important qualitative and quantitative discrepancies
are noted for the model by \citet{simonin93} when $10<y^+<30$. In the case of the highest particle inertia, this latter model and the present
one give surprisingly similar results which are in acceptable agreement with the DNS data while the predictions by the simple model with $D_i=0$
are poor. It could have been expected that this latter model would lead to better results for these particles. Nonetheless, the inertia of the
particle studied is not enough high to show that the present model and the one with $D_i=0$ should predict the drift velocity more accurately
than the model by \citet{simonin93}. The capability of the three models can be better distinguished from the results of the wall-normal
component of the drift velocity. The stochastic model with $D_i=0$ predicts a null drift velocity whatever the distance to the wall and the
particle inertia. It is in complete disagreement with the DNS data. The results obtained with the model by \citet{simonin93} are in satisfactory
accordance. Nevertheless, important discrepancies begin to arise as the particle inertia increases. Contrary to these two models, the expression
proposed to estimate the drift vector of the stochastic equation leads to a good estimation of this drift velocity whatever the particle
inertia. These observations can explain the predictions of the particle concentration by the three models making use of simple physical
considerations. The wall-normal drift velocity given by the stochastic model with $D_i=0$ is null. It has been also shown that this model
predicts an increase of particle concentration in the near-wall region whatever the particle inertia. This non-physical increase in the case of
the lowest particle inertia is induced by the fact that there is no mean force to counteract the accumulation of the particles in low-turbulence
regions (the so-called turbophoresis effect\cite{catatrvi,Reeks}). To explain the uniform concentration obtained with the model by
\citet{simonin93} whereas an increase should be observed near the wall in the case of the $\tau_p^+=5\text{ and }25$ particles, similar
considerations can be put forward. The predicted drift velocity given by this model, which is higher than that of the present model, is
certainly too high to make possible the accumulation of these particles.\\

\begin{figure}[!htb]
 \centering\includegraphics[width=\myfigwidth]{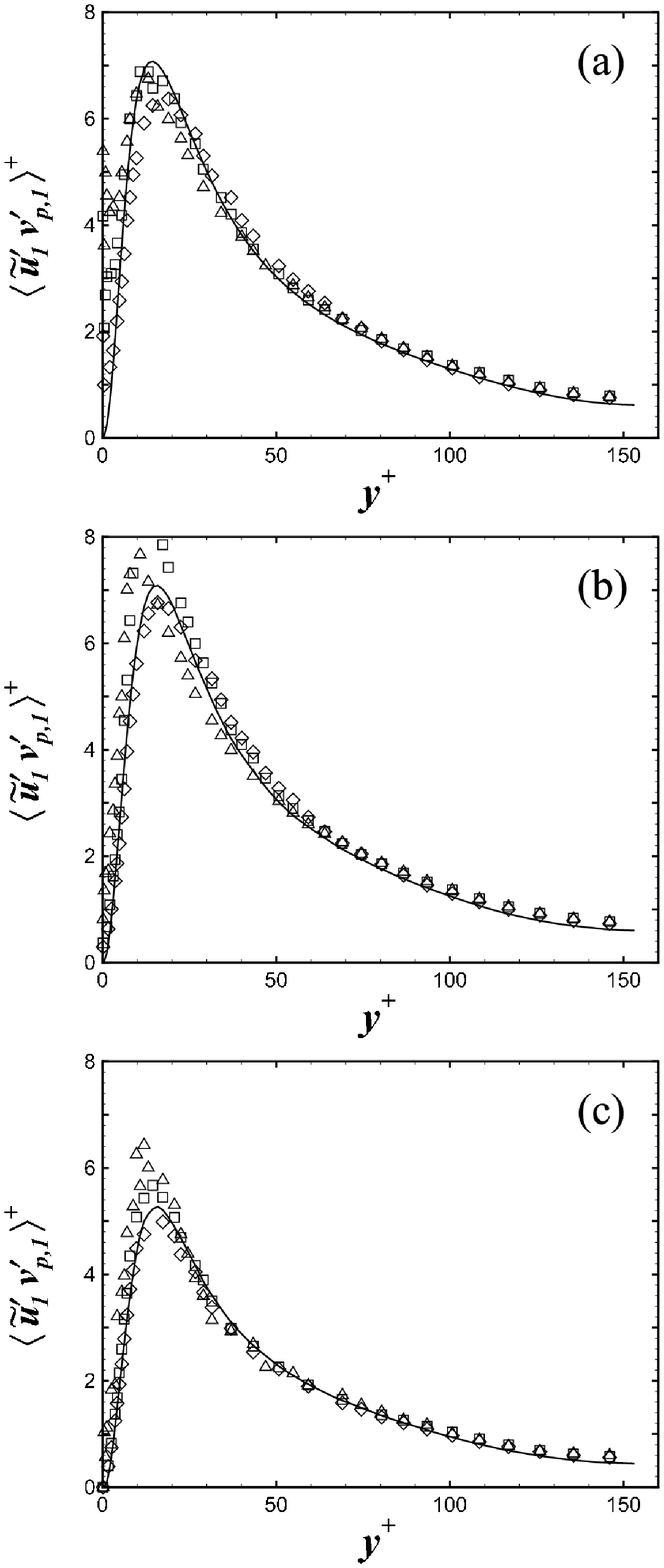}
 \caption{Diagonal component of the fluid-particle covariance tensor, $\lrc{\u_vu'_1v'_{p,1}}$, for $\tau_p^+=1$ (a), $\tau_p^+=5$ (b), and
 $\tau_p^+=25$ (c). DNS: --- . Stochastic simulation with $D_i=\partial\lrc{\u_vu'_iv'_{p,k}}/\partial x_k$ ($\Box$) ;
 $D_i=\partial\lrc{u'_iu'_k}/\partial x_k$ ($\triangle$) ; $D_i=0$ ($\Diamond$).}
\label{fig:vitesse_cov_part_11}
\end{figure}

\begin{figure}[!htb]
 \centering\includegraphics[width=\myfigwidth]{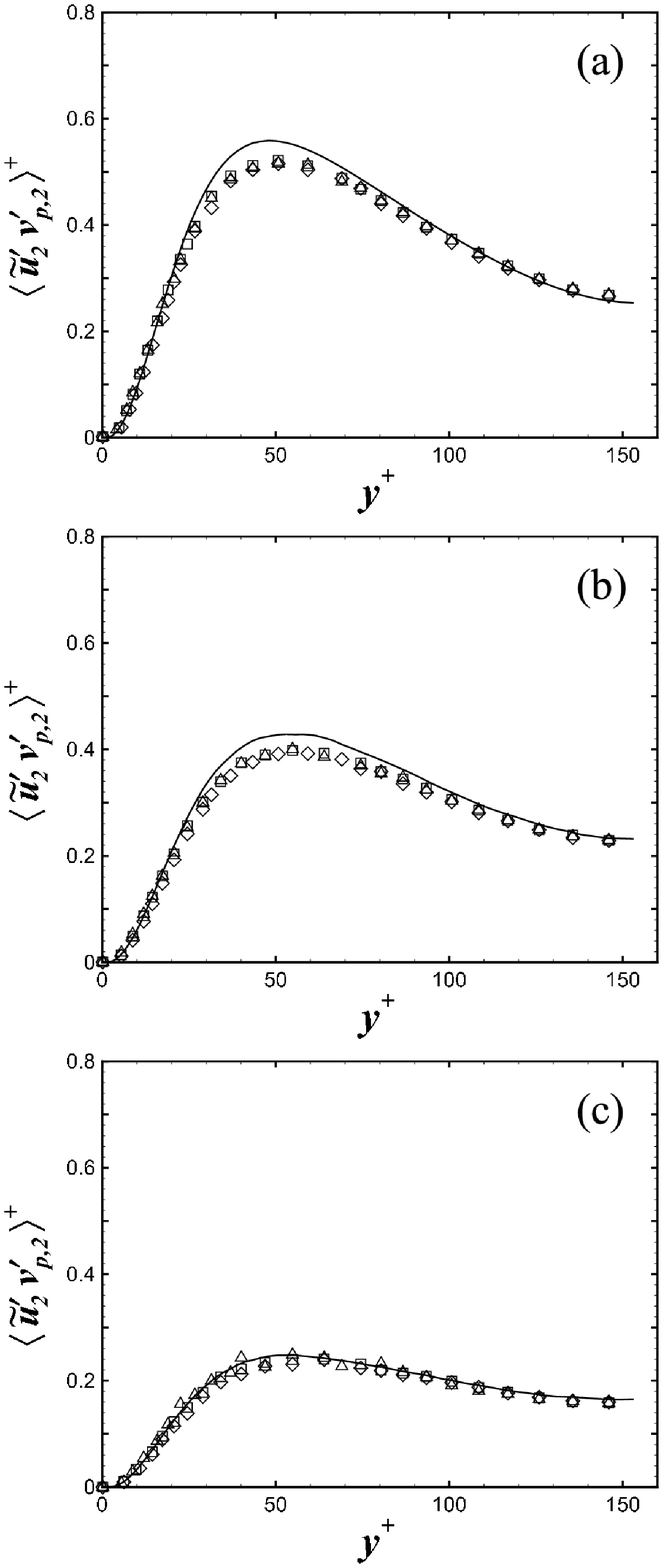}
 \caption{Diagonal component of the fluid-particle covariance tensor, $\lrc{\u_vu'_2v'_{p,2}}$, for $\tau_p^+=1$ (a), $\tau_p^+=5$ (b), and
 $\tau_p^+=25$ (c). DNS: --- . Stochastic simulation with $D_i=\partial\lrc{\u_vu'_iv'_{p,k}}/\partial x_k$ ($\Box$) ;
 $D_i=\partial\lrc{u'_iu'_k}/\partial x_k$ ($\triangle$) ; $D_i=0$ ($\Diamond$).}
\label{fig:vitesse_cov_part_22}
\end{figure}

\begin{figure}[!htb]
 \centering\includegraphics[width=\myfigwidth]{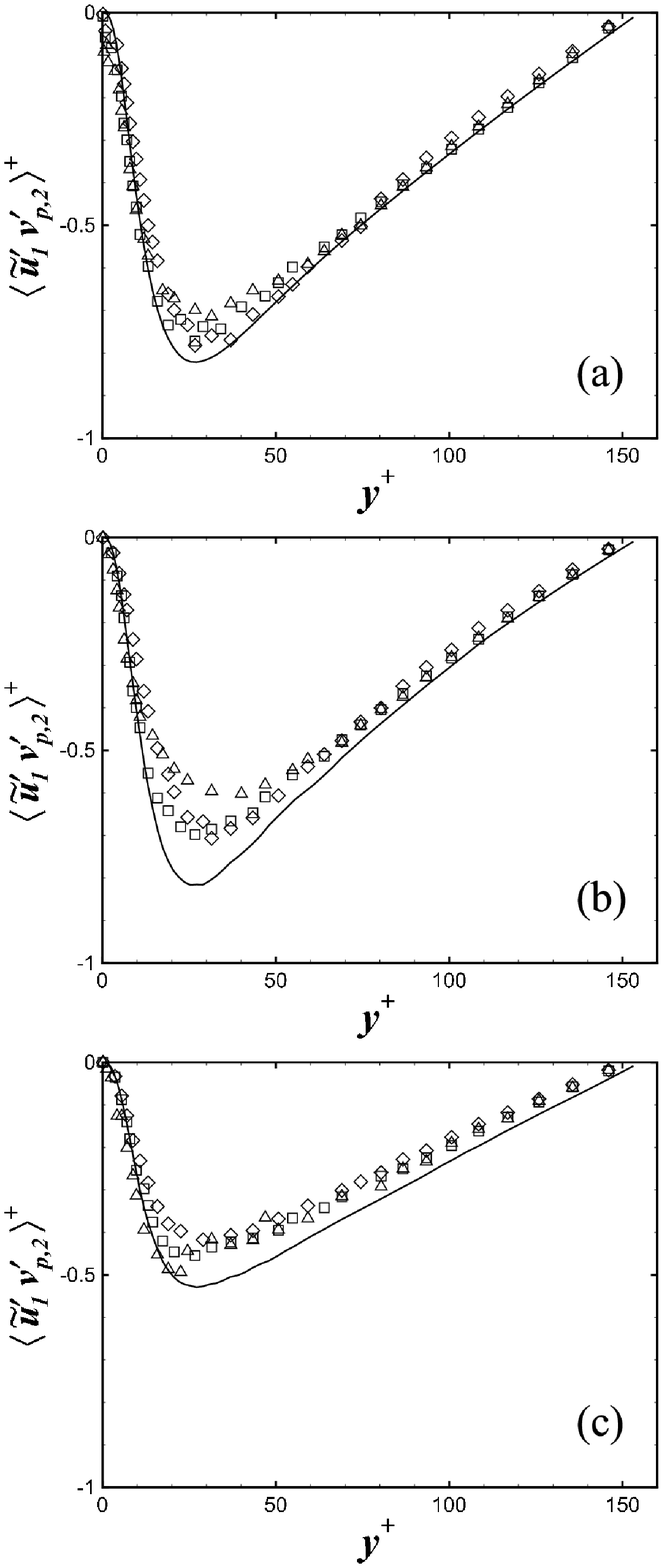}
 \caption{Non-diagonal component of the fluid-particle covariance tensor, $\lrc{\u_vu'_1v'_{p,2}}$, for $\tau_p^+=1$ (a), $\tau_p^+=5$ (b), and
 $\tau_p^+=25$ (c). DNS: --- . Stochastic simulation with $D_i=\partial\lrc{\u_vu'_iv'_{p,k}}/\partial x_k$ ($\Box$) ;
 $D_i=\partial\lrc{u'_iu'_k}/\partial x_k$ ($\triangle$) ; $D_i=0$ ($\Diamond$).}
\label{fig:vitesse_cov_part_12}
\end{figure}

\begin{figure}[!htb]
 \centering\includegraphics[width=\myfigwidth]{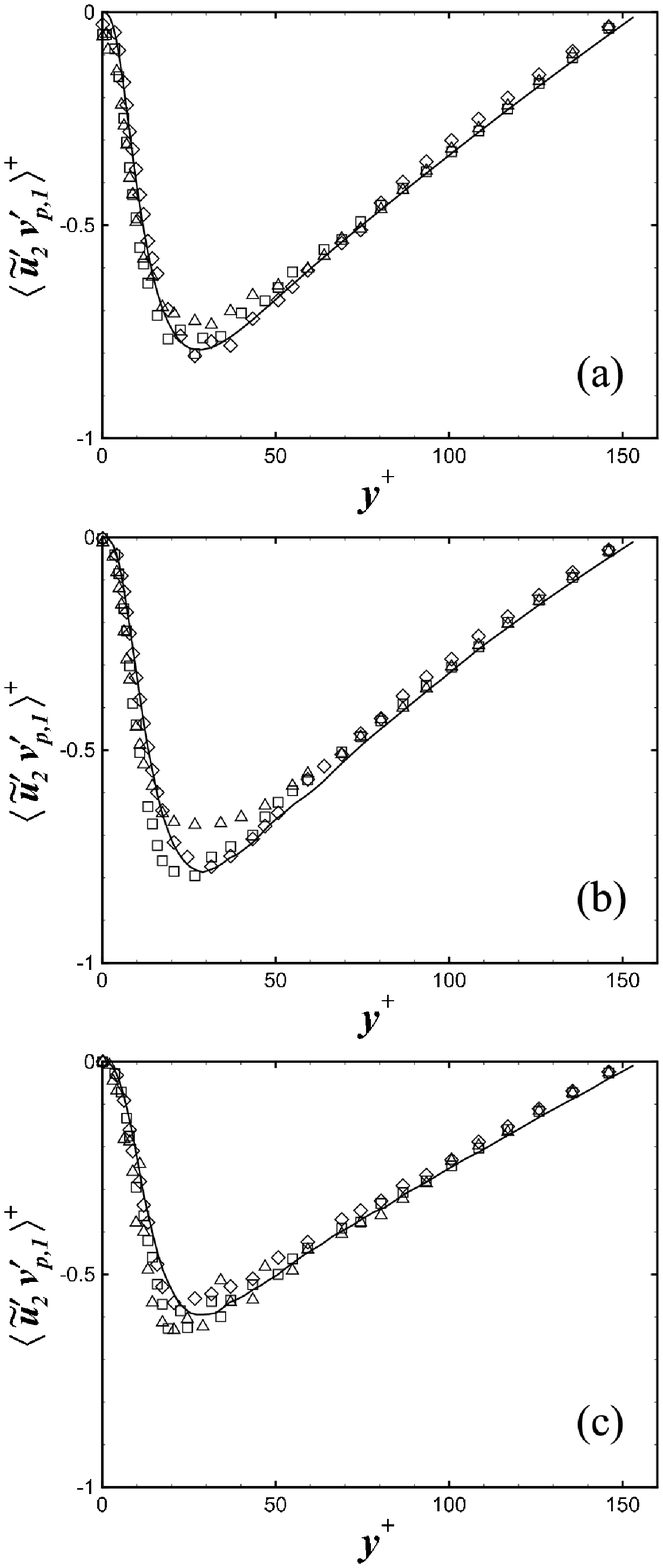}
 \caption{Non-diagonal component of the fluid-particle covariance tensor, $\lrc{\u_vu'_2v'_{p,1}}$, for $\tau_p^+=1$ (a), $\tau_p^+=5$ (b), and
 $\tau_p^+=25$ (c). DNS: --- . Stochastic simulation with $D_i=\partial\lrc{\u_vu'_iv'_{p,k}}/\partial x_k$ ($\Box$) ;
 $D_i=\partial\lrc{u'_iu'_k}/\partial x_k$ ($\triangle$) ; $D_i=0$ ($\Diamond$).}
\label{fig:vitesse_cov_part_21}
\end{figure}

The last statistical moment considered in this evaluation of the proposed stochastic model is the fluid seen-particle velocity correlations
$\lrc{\u_vu'_iv'_{p,j}}$. Concerning the diagonal components presented in Figs.~\ref{fig:vitesse_cov_part_11} and \ref{fig:vitesse_cov_part_22}, the agreement of the
stochastic simulations with the DNS data exhibits the same trends as for the particle velocity rms. The three models almost perfectly reproduce
the inertia effect on the wall-normal and spanwise components. It should be noted that due to the strong similarity between these two
components, only the results obtained for $\lrc{\u_vu'_2v'_{p,2}}$ are shown (see~Fig.~\ref{fig:vitesse_cov_part_22}). From
figure~\ref{fig:vitesse_cov_part_11}, it is noted that the streamwise component is in good accordance with the DNS data in the case of the
smaller particle inertia. Differences appear for higher particle inertia near the location of the maximum of $\lrc{\u_vu'_1v'_{p,1}}$
($10<y^+<20$). As observed for the particle velocity rms, the present model as well as the one by \citet{simonin93} lead to an overprediction of
the streamwise component while a better agreement is obtained with the third model [\eref{eq:GLM_fluc_general} with $D_i=0$]. Due to the
asymmetry of the fluid seen-particle velocity correlations, the two non-diagonal components are presented in
figures~\ref{fig:vitesse_cov_part_12} and \ref{fig:vitesse_cov_part_21}. The evolution of $\lrc{\u_vu'_1v'_{p,2}}$ as a function of the particle
inertia is well reproduced. Nonetheless, the stochastic models generally underestimate the DNS results. A better agreement is obtained for the
other non-diagonal component. As noted for the majority of the statistics presented in this study, the model by \citet{simonin93} provides an
acceptable but less accurate prediction.
%
%
\section{Conclusion}
We present in this study a stochastic model for estimating the fluid velocity experienced by small solid particles in a non-homogeneous
turbulent flow. In the first part, a new stochastic model which is compatible with the limits of the transport equation of the drift velocity
for low and high particle inertia has been derived. From this compatibility criterion, it has also been shown that some previously proposed
stochastic models should not be able to reflect accurately the inertia effect on the fluid velocity seen by high inertia particles.

In the second part of this study, the accuracy of the present stochastic equation has been evaluated. Since no models exist to determine the
drift and diffusion parameters, appearing in the SDE, for non-homogeneous turbulence, they have been deduced from a method similar to the one
proposed by \citet{pope_02}. The obtained results show that the drift and diffusion matrices are highly space-dependent and anisotropic. Using
these values, stochastic simulations of a gas-solid channel flow have been conducted and compared to DNS. The stochastically predicted data have
been also compared to those obtained with the model proposed by \citet{simonin93} and to a simpler one. Using the compatibility criterion
presented in the first part, the former model should not be able to reproduce the dynamics of high particle inertia while the latter should not
be able to reproduce it for small particle inertia.\\
The three models considered are able to predict with a good accuracy the first and second order statistical moments of the particle and fluid
seen velocities. Surprisingly, a good accordance has been also noticed for the triple particle velocity correlations. This accordance is mainly
qualitative. Nevertheless, some components have been seen to be well predicted quantitatively. This clearly demonstrates the
capability of Langevin-type models to predict accurately and efficiently the interactions between inertial particles and turbulence.\\
The accuracy of the results obtained with these three different models diverges principally for the particle concentration and the drift
velocity. As stated before, a good estimation of these quantities is primordial to correctly predict the important process of particle
deposition. It has been seen that the model proposed by \citet{simonin93} is not able to predict the increase near the wall of the concentration
of moderate and high particle inertia. On contrary, the simpler model predicts this increase even for the smaller particle inertia whereas the
concentration should be almost uniform. This clearly shows that this model suffers from a spurious drift effect. The new proposed model is the
only one which succeeds in predicting the good evolution of the particle concentration for the range of particle inertia studied. This naturally
leads us to consider that it is a good candidate to estimate the turbulence seen by inertial particles. In the present paper, we test the
proposed model using DNS data such as for the fluid-particle covariances in order to not introduce supplementary modeling uncertainties. These
data are generally not known beforehand. Nonetheless, numerical strategies related to RANS-Lagrangian methods can help to overcome this
difficulty. For instance, one such method can be found in \citet{pechpomi}, which calculates the fluid-particle covariances on the basis of the
statistics of a large number of particles. Thus, the model is more computationally demanding than Simonin's method, but it is hoped
that the benefit of better results outweighs the extra cost.\\
Another possible way to bypass this difficulty would be to directly model the fluid-particle covariances. The simplest existing model is based
on the theory developed by \citet{tchen} and \citet{hinze} [see \citet{simonin93}]. Attention must be paid to the properties of the carrier
fluid flow studied since this model was initially developed for isotropic and stationary turbulence under restrictive assumptions. A more
sophisticated approach was recently proposed  by \citet{zaoeal}. Although their model was developed for quasi-homogeneous anistropic turbulence,
quite accurate predictions of the fluid-particle covariances were obtain for a gas-solid non-homogeneous flow.\cite{artaza2} Moreover, the union
of these models with the proposed stochastic equation is consistent since both depend on the same set of quantities, i.e. the decorrelation time
scales and second order statistical moment of the fluid velocity seen by particles.

We would like to emphasize that the model proposed belongs to a particular class of stochastic models. Many other models could certainly
reproduce more accurately the interactions between inertial particles and turbulence. Nonetheless, the difficulty is to find a model which is a
good compromise between complexity and physical accuracy. Besides, the challenge lies also in the specification of the model parameters. One
could propose a model which is theoretically able to reproduce the different physical aspects of gas-solid flows with high fidelity, but if its
parameters cannot be estimated accurately, the model will probably be less satisfactory than a simpler model whose parameters can be found
precisely. One part of the physics is in the functional form of the model, the other part is in its parameters.

Finally, it should be noted that Lagrangian stochastic methods can also be considered for predicting the motion of gas bubbles in a liquid.
Nevertheless, deformation, coalescence and break-up can significantly modify bubbles shape, and consequently, alter the interactions of each
bubbles with turbulence. This makes the use of a stochastic model more complex since its parameters, which are not well known, should be
correctly modified during the bubble tracking. In addition, the use of the point-force approximation could become ambiguous when bubble
coalescence is strong. However, if we restrict ourselves to small non-deformable bubbles and neglect break-up and coalescence, a Lagrangian
stochastic method can be retained. As for solid particles, the difficulty will be to properly estimate the parameters of the stochastic model.%

%

\newpage
\listoffigures


\end{document}